\documentclass[superscriptaddress,twocolumn,amsmath,amssymb,prd,floatfix,nofootinbib]{revtex4-2}

\usepackage[dvipdfmx]{graphicx}
\usepackage{dcolumn}
\usepackage[varg]{txfonts}
\usepackage{bm}
\usepackage{multirow}

\usepackage{color}

\newcommand*\Laplace{\mathop{}\!\mathbin\bigtriangleup}

\begin{document}

\title{Integrated perturbation theory for cosmological tensor
  fields. II. Loop corrections
}

\author{Takahiko Matsubara} \email{tmats@post.kek.jp}
\affiliation{%
  Institute of Particle and Nuclear Studies, High Energy
  Accelerator Research Organization (KEK), Oho 1-1, Tsukuba 305-0801,
  Japan}%
\affiliation{%
  The Graduate Institute for Advanced Studies, SOKENDAI,
  Tsukuba 305-0801, Japan}%

\date{\today}

\begin{abstract}
  In the previous paper \cite{PaperI}, the nonlinear perturbation
  theory of the cosmological density field is generalized to include
  the tensor-valued bias of astronomical objects, such as spins and
  shapes of galaxies and any other tensors of arbitrary ranks which
  are associated with objects that we can observe. We apply this newly
  developed method to explicitly calculate nonlinear power spectra and
  correlation functions both in real space and in redshift space.
  Multidimensional integrals that appear in loop corrections are
  reduced to combinations of one-dimensional Hankel transforms, thanks
  to the spherical basis of the formalism, and the final expressions
  are numerically evaluated in a very short time using an algorithm of
  the fast Fourier transforms such as \textsc{FFTLog}. As an
  illustrative example, numerical evaluations of loop corrections of
  the power spectrum and correlation function of the rank-2 tensor
  field are demonstrated with a simple model of tensor bias.
\end{abstract}


\maketitle


\section{Introduction
\label{sec:Introduction}}

The large-scale structure (LSS) of the Universe, probed by galaxies
and other astronomical observables such as weak lensing, 21~cm
emission and absorption lines and so forth, plays an essential role in
cosmology. The LSS is complementary to the cosmic microwave background
(CMB) radiation, which mainly probes the early stages of the Universe
around the time of decoupling. The information contained in the
temperature fluctuations in the CMB have been extracted in exquisite
details, and the temperature and polarization maps obtained by the
Planck satellite determined the precise values of the cosmological
parameters \cite{Planck:2018vyg}. Beyond the cosmological information
extracted from the CMB, the LSS offers a lot of opportunities to
obtain further information of the Universe which is contained mainly
in the lower-redshift Universe. In addition, the representative values of
the cosmological parameters determined by the Planck are obtained by
combining the observational data of LSS, such as the scale of baryon
acoustic oscillations (BAO) and weak lensing, as the CMB data alone
has degeneracies among cosmological parameters. The acceleration of
the Universe due to dark energy is also an effect that can only be
probed in the lower-redshift Universe.

While most of the physics in CMB is captured by the linear
perturbation theory of fluctuations, the properties of LSS are more
affected by nonlinear evolutions, as the scales of interest become
smaller. On one hand, the physics of long-wavelength modes in the
density fluctuations in the LSS can still be captured by the linear
perturbation theory, and the amplitude of density fluctuations simply
grows according to the linear growth factor. However, the number of
independent modes of density fluctuations included in a survey is
limited by the finiteness of the survey volume $V$, i.e., the number
of independent Fourier modes with a wave number magnitude $k$ roughly
scales as $\sim k^3V$ in three-dimensional surveys. On the other hand,
short-wavelength modes are affected by nonlinear evolutions of the
density field, which mix up different scales of Fourier modes, and
thus their analysis becomes much more difficult. The fully nonlinear
evolutions cannot be analytically solved because of the extreme
mixture of modes, and extraction of the cosmological information from
the fully nonlinear density field is difficult. While one can resort
to the numerical simulations to solve the nonlinear evolutions, the
information contents of initial condition and cosmology are largely
lost in the nonlinearly evolved field \cite{Rimes:2005xs}, compared to
the linearly evolved field.

The transition scales of the linear and nonlinear fields are roughly
around $10$--$20\,h^{-1}\mathrm{Mpc}$ at the present time of $z=0$,
and the transition scales become smaller at an earlier time of higher
redshift. On the transition scales, although the linear theory does
not quantitatively apply, the nonlinearity is still weak and the
mixing of Fourier modes is not complicated enough. Only a countable
number of modes are effectively mixed, and the nonlinear perturbation
theory is applicable in such a situation. Therefore, the theory of
nonlinear perturbation theory of density field
\cite{Bernardeau:2008fa} is expected to play an important role in the
analysis of the LSS, in the era of large surveys in the near future when
the sufficiently large number of Fourier modes on the transition
scales are expected to be available. In addition, the density
fluctuations even on large scales, which have been traditionally
considered as the linear regime, are more or less affected by weak
nonlinearity, and it is critically important to estimate such subtle
effects in the era of precision cosmology. A representative example of
the last case is the nonlinear smearing effects of the BAO in
correlation functions of galaxies around
$\sim 100\,h^{-1}\mathrm{Mpc}$ \cite{SDSS:2005xqv}, which is used as a
powerful standard ruler to probe the expansion history of the Universe
and the nature of dark energy.

The higher-order perturbation theory beyond the linear theory has
been extensively developed for matter distributions in the past several
decades
\cite{Jusz1981,Vish1983,Fry:1983cj,Goroff:1986ep,Suto:1990wf,Makino:1991rp,Jain:1993jh,Jeong:2006xd}.
However, the distribution of matter is not the same as that of
objects that we can observe, and the mass density field is dominated
by the dark matter in the Universe. The bias between distributions of
matter and observable objects is one of the most important concepts in
understanding the large-scale structure of the Universe. In order that
the nonlinear perturbation theory can be compared with observations,
the effect of bias is an indispensable element that should be included
in the theories with the predictability of the observable Universe. There
are many attempts to include the effect of bias in the nonlinear
perturbation theory (for a recent review, see
Ref.~\cite{Desjacques:2016bnm}). Understanding the bias from the first
principle is extremely difficult because of the full nonlinearity of
the problem and extremely complicated astrophysical processes in the
galaxy formation, and so forth.

The concept of bias has usually been discussed in the context of
number density fields of astronomical objects such as galaxies, as
probes of the underlying matter density field. In this case, the bias
corresponds to a function, or more properly a functional, of the
underlying mass density field to give a number density field of the
biased objects. Thus the function(al) of the bias has a scalar value
in accordance with that the density of biased objects is a scalar
field. In the previous work of Paper I \cite{PaperI}, the concept of
bias in the nonlinear perturbation theory is generalized to the case
that the bias is given to a tensor field. The number densities of
objects are not the only probes of the density fields in the LSS. For
example, galaxy spins and shapes are in principle determined by the
mass density fields through, e.g., tidal gravitational forces, and
other physical quantities. Recently, interests in statistics of the
galaxy sizes and shapes, or intrinsic alignments, are growing as
probes of the LSS of the Universe
\cite{Catelan:2000vm,Okumura:2008du,Joachimi:2015mma,Kogai:2018nse,Chisari:2013dda},
and analytical modelings of galaxy shape statistics by the nonlinear
perturbation theory have also been introduced
\cite{Blazek:2015lfa,Blazek:2017wbz,Schmitz:2018rfw,Vlah:2019byq,Vlah:2020ovg}.

Motivated by these recent developments, we generalize the nonlinear
perturbation theory to predict statistics of biased fields with an
arbitrary rank of tensor in Paper I \cite{PaperI}. We adopt the
spherical decomposition of the tensor field, which plays an important
role in the formalism. This method of decomposition has been already
adopted in the perturbation theory in literature to investigate the
clustering of density peaks \cite{Desjacques:2010gz} and galaxy shapes
\cite{Vlah:2019byq}. In the last two references, the coordinates
system of the spherical basis is chosen so that the third axis is
aligned with a radial direction of the correlation function, or a
direction of wave vector of perturbations in Fourier space. In
contrast, we do not fix the coordinates system in the spherical basis,
and explicitly keep the rotational covariance apparent throughout the
formulation. The basic formalism to calculate the power spectrum and
higher-order spectra of tensor fields of arbitrary ranks by the
nonlinear perturbation theory to arbitrary orders is described in
Paper~I.

Many different methods have been considered in the literature to
include the bias in the nonlinear perturbation theory
\cite{Desjacques:2016bnm}. Most methods rely on a local or semilocal
ansatz of the bias function which relates the mass density field and
the biased density field. The locality or semilocality of the relation
is given in either Eulerian or Lagrangian space of the density field.
However, (semi)local biases in Eulerian and Lagrangian spaces are not
compatible with each other in general, because the dynamically
nonlinear evolution by gravity is essentially nonlocal. Therefore, the
bias relation should be given by a nonlocal functional, in either
Eulerian or Lagrangian space, and the (semi)local \textit{Ans\"atze}
are approximations to the reality. A general formulation to
systematically incorporate the nonlocal bias into the nonlinear
perturbation theory is provided by the integrated perturbation theory
(iPT) \cite{Matsubara:2011ck,Matsubara:2013ofa}. The local and
semilocal \textit{Ansatz} of the bias can also be derived from this
formulation by restricting the form of bias functional in the class of
local or semilocal function. Moreover, the iPT also provides a
natural way of including the effect of redshift space distortions,
which should be taken into account for predicting observable
statistics in redshift surveys. Our formulation of Paper~I is built
upon and generalizes the method of iPT and establishes a nonlinear
perturbation theory of tensor fields in general. Paper~I describes the
basic formulation of the theory and gives some results of lowest-order
approximations of the perturbation theory.

In this second paper of the series, we apply the formulation of
Paper~I to concretely calculate the one-loop corrections of the
perturbation theory. The strategy of the calculation is fairly
straightforward according to Paper~I. Some techniques are introduced
to reduce the higher-dimensional integrals to the lower ones, which
are generalizations of an existing method using a fast Fourier
transform applied to the nonlinear perturbation theory
\cite{Schmittfull:2016jsw}. In particular, all the necessary
integrations to evaluate the one-loop corrections in the perturbation
theory with the (semi)local models of tensor bias reduce to
essentially one-dimensional Hankel transforms. As an illustrative
example, we calculate the power spectrum and correlation function with
one-loop corrections for a simple model of a rank-2 tensor which is
biased from spatially second derivatives of the gravitational
potential in Lagrangian space.

This paper is organized as follows. In Sec.~\ref{sec:Propagators}, the
propagators, elements of the nonlinear perturbation theory, in the
spherical basis of our formalism are calculated, up to necessary
orders for evaluating one-loop corrections of the power spectrum and
correlation function. In Sec.~\ref{sec:OneLoopPS}, our main result,
the one-loop approximations of the power spectra of the tensor field
are explicitly derived in analytic forms, both in real space and in
redshift space. In Sec.~\ref{sec:SimpleEx}, a simple example of the
tensor bias with a semilocal model is explicitly calculated and
numerically evaluated. Conclusions are given in
Sec.~\ref{sec:Conclusions}. In the Appendix, a formal expression of
the all-order power spectrum of the tensor field is derived beyond the
one-loop approximation.

\section{\label{sec:Propagators}
  Propagators of tensor fields and loop corrections
}

The fundamental formulation of the iPT of tensor fields is described
in Paper~I \cite{PaperI}. One of the essential ingredients of the
theory is the evaluation of propagators, with which statistics of
tensor fields, such as the power spectrum, bispectrum, correlation
functions, etc.~are represented. Several examples in relatively simple
cases with lowest-order approximation are explicitly derived in Sec.~V
of Paper~I. In this section, we further derive the propagators that
are required to evaluate next-order approximation with loop
corrections. We cite many equations from Paper~I, which readers are
assumed to have in hand.

\subsection{\label{subsec:InvProp}
  Invariant propagators
}

The propagators of tensor fields can be represented by
rotationally invariant functions as well as the renormalized bias
functions as extensively explained in Paper~I. First, we summarize the
essential equations and introduce various quantities and functions
which are used in later sections. The concepts of propagators and
renormalized bias functions are explained in detail in Sec.~III of
Paper~I. The details of the definitions are not explained here. In
short, they are response functions of the nonlinear evolutions from
the initial density field. Below we summarize their properties which
are essential to derive the main equations in later sections of this
paper.

\subsubsection{\label{subsubsec:InvPropReal}
  Real space
}

The reduced propagators (see Sec.~III~A of Paper~I for their
definitions) up to the second order are represented by invariant
functions as
\begin{equation}
  \hat{\Gamma}^{(1)}_{Xlm}(\bm{k}) =
  \frac{(-1)^l}{\sqrt{2l+1}}\,
  \hat{\Gamma}^{(1)}_{Xl}(k) C_{lm}(\hat{\bm{k}}),
  \label{eq:1}
\end{equation}
for the first order, and
\begin{equation}
  \hat{\Gamma}^{(2)}_{Xlm}(\bm{k}_1,\bm{k}_2)
  = \sum_{l_1,l_2}
  \hat{\Gamma}^{(2)\,l}_{Xl_1l_2}(k_1,k_2)
  X^{l_1l_2}_{lm}(\hat{\bm{k}}_1,\hat{\bm{k}}_2),
  \label{eq:2}
\end{equation}
for the second order. In the above equations, the index $X$ specifies
the class of tensor-valued objects in general, such as the density
(scalar), angular momentum (vector) or shape (tensors) of a certain type
of galaxies etc. The functions $\hat{\Gamma}^{(1)}_{Xl}(k)$ and
$\hat{\Gamma}^{(2)\,l}_{Xl_1l_2}(k_1,k_2)$ are the invariant
propagators which are invariant under the coordinates rotations. We
use the spherical harmonics with Racah's normalization,
\begin{equation}
  C_{lm}(\theta,\phi)
  \equiv \sqrt{\frac{4\pi}{2l+1}} Y_{lm}(\theta,\phi)
  = \sqrt{\frac{(l-m)!}{(l+m)!}}
  P_l^m(\cos\theta)\,e^{im\phi},
  \label{eq:3}
\end{equation}
instead of standard normalization of spherical harmonics $Y_{lm}$. The
arguments of the spherical harmonics are alternatively represented by
a unit vector $\bm{n}$, instead of the corresponding angular
coordinates $(\theta,\phi)$ of $\bm{n}$. In the above notation, the
Condon-Shortley phase is included in the associated Legendre
polynomials $P_l^m$ as
\begin{equation}
  P_l^m(x) =
  \frac{(-1)^m}{2^l\,l!}
  \left( 1-x^2 \right)^{m/2}
  \frac{d^{l+m}}{dx^{l+m}}
  \left( 1-x^2 \right)^l.
  \label{eq:4}
\end{equation}
The function of the last factor in Eq.~(\ref{eq:2}) is the bipolar
spherical harmonics with an appropriate normalization,
\begin{equation}
  X^{l_1l_2}_{lm}(\bm{n}_1,\bm{n}_2)
  = \left(l\,l_1\,l_2\right)_{m}^{\phantom{m}m_1m_2}
    C_{l_1m_1}(\bm{n}_1) C_{l_2m_2}(\bm{n}_2),
  \label{eq:5}
\end{equation}
where azimuthal indices $m_1$ and $m_2$ are summed over from $-l_1$ to
$+l_1$ and from $-l_2$ to $+l_2$, respectively, without summation
symbols following the Einstein convention, and
\begin{equation}
  \left(l\,l_1\,l_2\right)_m^{\phantom{m}m_1m_2}
  = (-1)^{m_1+m_2}
  \begin{pmatrix}
    l & l_1 & l_2 \\
    m & -m_1 & -m_2
  \end{pmatrix}
  \label{eq:6}
\end{equation}
is a Wigner's $3j$-symbol.

It is convenient to use the metric tensor of spherical basis, defined
by
\begin{equation}
  g_{(l)}^{mm'} = g^{(l)}_{mm'} = (-1)^m \delta_{m,-m'},
  \label{eq:7}
\end{equation}
where $\delta_{m,-m'}$ is the Kronecker's symbol which is unity when
$m+m'=0$ and is zero otherwise. With this
notation, Eq.~(\ref{eq:6}) is represented by
\begin{equation}
  \left(l\,l_1\,l_2\right)_{m}^{\phantom{m}m_1m_2} =
  g_{(l)}^{m_1m_1'} g_{(l)}^{m_2m_2'}
  \left(l\,l_2\,l_3\right)_{mm_1'm_2'},
\label{eq:8}
\end{equation}
where 
\begin{equation}
  \left(l_1\,l_2\,l_3\right)_{m_1m_2m_3} =
  \begin{pmatrix}
    l_1 & l_2 & l_3 \\
    m_1 & m_2 & m_3
  \end{pmatrix}
\label{eq:9}
\end{equation}
is the usual $3j$-symbol, and Einstein's summation convention for the
azimuthal indices $m$, $m_1$, etc.~are assumed throughout this paper,
unless otherwise stated. The two spherical metric tensors satisfy
$g_{(l)}^{mm''} g^{(l)}_{m''m'} = \delta^m_{m'}$. Similarly to
Eq.~(\ref{eq:8}), we understand that the azimuthal indices can be
raised or lowered by the spherical metric tensor, for example,
\begin{equation}
  \left(l_1\,l_2\,l_3\right)_{m_1m_2}^{\phantom{m_1m_2}m_3} =
  g_{(l)}^{m_3m_3'} \left(l_1\,l_2\,l_3\right)_{m_1m_2m_3'},
  \label{eq:10}
\end{equation}
and so forth.

With the above notation, the complex conjugate of the spherical
harmonics is represented by
\begin{equation}
  C^*_{lm}(\bm{n}) = g_{(l)}^{mm'} C_{lm'}(\bm{n}),
  \label{eq:11}
\end{equation}
and similarly, that of the bipolar spherical harmonics is represented
by
\begin{equation}
  X^{l_1l_2*}_{lm}(\hat{\bm{k}}_1,\hat{\bm{k}}_2)
  = (-1)^{l_1+l_2+l} g_{(l)}^{mm'}
  X^{l_1l_2}_{lm'}(\hat{\bm{k}}_1,\hat{\bm{k}}_2).
  \label{eq:12}
\end{equation}

The orthonormality relation of spherical harmonics is given by
\begin{equation}
  \int \frac{d^2n}{4\pi}\,
  C_{lm}^*(\bm{n}) C_{l'm'}(\bm{n}) =
  \frac{\delta_{ll'}}{2l+1} \delta^m_{m'},
  \label{eq:13}
\end{equation}
and those of bipolar spherical harmonics is given by
\begin{multline}
  \int \frac{d^2\hat{k}_1}{4\pi} \frac{d^2\hat{k}_2}{4\pi}
  X^{l_1l_2*}_{lm}(\hat{\bm{k}}_1,\hat{\bm{k}}_2)
  X^{l_1'l_2'}_{l'm'}(\hat{\bm{k}}_1,\hat{\bm{k}}_2)
  \\
  = \frac{\delta_{ll'} \delta_{l_1l_1'} \delta_{l_2l_2'}
    \delta^\triangle_{l_1l_2l}}{(2l+1)(2l_1+1)(2l_2+1)} 
  \delta^m_{m'},
  \label{eq:14}
\end{multline}
where $\delta^\triangle_{l_1l_2l}$ is unity when the set of integers
$(l_1,l_2,l)$ satisfies triangle inequality $|l_1-l_2| \leq l \leq
l_1+l_2$, and is zero otherwise.

The invariant propagator of the second order satisfies an interchange
symmetry,
\begin{equation}
  \hat{\Gamma}^{(2)\,l}_{Xl_2l_1}(k_2,k_1)
  = (-1)^{l_1+l_2+l}\,
  \hat{\Gamma}^{(2)\,l}_{Xl_1l_2}(k_1,k_2).
  \label{eq:15}
\end{equation}
The above expansions of Eqs.~(\ref{eq:1}) and (\ref{eq:2}) are
inverted by the above orthonormality relations, and we have
\begin{equation}
  \hat{\Gamma}^{(1)}_{Xl}(k) =
  (-1)^l\sqrt{2l+1}\, g_{(l)}^{mm'}
  \int \frac{d^2\hat{k}}{4\pi}\,
  \hat{\Gamma}^{(1)}_{Xlm}(\bm{k})
  C_{lm'}(\hat{\bm{k}})
  \label{eq:16}
\end{equation}
for the first-order propagator, and
\begin{multline}
  \hat{\Gamma}^{(2)\,l}_{Xl_1l_2}(k_1,k_2)
  = (2l_1+1)(2l_2+1)\, g_{(l)}^{mm'}
  \\ \times
  \int \frac{d^2\hat{k}_1}{4\pi} \frac{ d^2\hat{k}_2}{4\pi}\,
  \hat{\Gamma}^{(2)}_{Xlm}(\bm{k}_1,\bm{k}_2)
  X^{l_1l_2}_{lm'}(\hat{\bm{k}}_1,\hat{\bm{k}}_2)
  \label{eq:17}
\end{multline}
for the second-order propagator. In practice, one can always represent the
propagators with polypolar spherical harmonics in the form of
Eqs.~(\ref{eq:1}) and (\ref{eq:2}), and can readily read off the
expression of the invariant functions from the results.

\subsubsection{\label{subsubsec:InvPropRed}
  Redshift space
}

In redshift space, the propagators also depend on the direction of the
line of sight. We can decompose the dependence on the line of sight in
spherical harmonics, together with the dependencies on the directions
of wave vectors. For the first-order propagator in redshift space, we
have
\begin{equation}
  \hat{\Gamma}^{(1)}_{Xlm}(\bm{k};\hat{\bm{z}};k,\mu)
  = \sum_{l_z,l_1}
  \hat{\Gamma}^{(1)\,l\,l_z}_{Xl_1}(k,\mu)
  X^{l_zl_1}_{lm}(\hat{\bm{z}},\hat{\bm{k}}),
  \label{eq:18}
\end{equation}
where $\hat{\bm{z}}$ is the direction to the line of sight. We assume
the distant-observer approximation, and the direction to the line of
sight is fixed in space. Unlike the common practice, we do {\em not}
fix the line of sight in the third direction of the coordinates, but
allow to point in any direction. In the above expression, the
direction cosine to the line of sight,
$\mu \equiv \hat{\bm{z}}\cdot\hat{\bm{k}}$, is included. This
dependence is not necessarily included there, because the angular
dependence on the left-hand side (lhs) of the equation can be
completely expanded into spherical harmonics (Sec.~IV~B~2 of Paper~I).
However, it is sometimes convenient to leave some part of the
dependence in the form of the direction cosine $\mu$ between the line
of sight and the direction of the wave vector. Which part of the dependence
is kept unexpanded is arbitrary. Even though the arguments $k$ and
$\mu$ of the propagator on the lhs of Eq.~(\ref{eq:18}) is a function
of $\bm{k}$ and $\hat{\bm{z}}$, the explicit arguments of $k$ and
$\mu$ specify which parts of the angular dependence in these
parameters are unexpanded in the spherical harmonics on the right-hand
side (rhs).

For the second-order propagator in redshift space, we have
\begin{multline}
  \hat{\Gamma}^{(2)}_{Xlm}(\bm{k}_1,\bm{k}_2;\hat{\bm{z}};k,\mu)
  \\
  =
  \sum_{l_z,l_1,l_2,L}
  \hat{\Gamma}^{(2)\,l\,l_z;L}_{Xl_1l_2}(k_1,k_2;k,\mu)
  X^{l_zl_1l_2}_{L;lm}(\hat{\bm{z}},\hat{\bm{k}}_1,\hat{\bm{k}}_2),
  \label{eq:19}
\end{multline}
where
\begin{multline}
  X^{l_1l_2l_3}_{L;lm}(\bm{n}_1,\bm{n}_2,\bm{n}_3)
  = (-1)^L \sqrt{2L+1}
  \left(l\,l_1\,L\right)_{m}^{\phantom{m}m_1M}
  \left(L\,l_2\,l_3\right)_{M}^{\phantom{M}m_2m_3}
  \\ \times
  C_{l_1m_1}(\bm{n}_1) C_{l_2m_2}(\bm{n}_2)
  C_{l_3m_3}(\bm{n}_3)
  \label{eq:20}
\end{multline}
is the tripolar spherical harmonics with an appropriate normalization.
In the argument of propagators, the variables $k=|\bm{k}|$ and
$\mu=\hat{\bm{z}}\cdot\hat{\bm{k}}$ are given by the total wave vector
$\bm{k}=\bm{k}_1+\bm{k}_2$, which is optionally allowed to be
included, because keeping the angular dependencies in these variables
significantly simplifies the analytic calculations. The invariant
propagator of the second order satisfies an interchange symmetry,
\begin{equation}
  \hat{\Gamma}^{(2)\,l\,l_z;L}_{Xl_2l_1}(k_2,k_1;k,\mu)
  = (-1)^{l_1+l_2+L}
  \hat{\Gamma}^{(2)\,l\,l_z;L}_{Xl_1l_2}(k_1,k_2;k,\mu).
  \label{eq:21}
\end{equation}
The complex conjugate of the tripolar spherical harmonics is given by
\begin{equation}
  X^{l_1l_2l_3*}_{lm}(\bm{n}_1,\bm{n}_2,\bm{n}_3)
  =(-1)^{l+l_1+l_2+l_3} g_{(l)}^{mm'}
  X^{l_1l_2l_3}_{lm'}(\bm{n}_1,\bm{n}_2,\bm{n}_3),
  \label{eq:22}
\end{equation}
and the orthonormality relation is given by
\begin{multline}
  \int \frac{d^2n_1}{4\pi} \frac{d^2n_2}{4\pi} \frac{d^2n_3}{4\pi}
  X^{l_1l_2l_3}_{L;lm}(\bm{n}_1,\bm{n}_2,\bm{n}_3)
  X^{l_1'l_2'l_3'}_{L';l'm'}(\bm{n}_1,\bm{n}_2,\bm{n}_3)
  \\
  = \frac{(-1)^{l+l_1+l_2+l_3} \delta_{ll'} \delta_{l_1l_1'} \delta_{l_2l_2'}
    \delta_{l_3l_3'} \delta_{LL'} \delta^\triangle_{l\,l_1L}
    \delta^\triangle_{Ll_2l_3}}{(2l+1)(2l_1+1)(2l_2+1)(2l_3+1)} 
    g^{(l)}_{mm'}.
  \label{eq:23}
\end{multline}

Applying the above orthonormality relations for bipolar and tripolar
spherical harmonics, Eqs.~(\ref{eq:14}) and (\ref{eq:23}), 
the first- and second-order propagators in redshift space
of Eqs.~(\ref{eq:18}) and (\ref{eq:19}) are inverted as
\begin{multline}
  \hat{\Gamma}^{(1)\,l\,l_z}_{Xl_1}(k,\mu)
  = (2l_z+1)(2l_1+1)\,g_{(l)}^{mm'}
  \\ \times
  \int \frac{d^2\hat{z}}{4\pi} \frac{d^2\hat{k}}{4\pi}\,
  \hat{\Gamma}^{(1)}_{Xlm}(\bm{k};\hat{\bm{z}};k,\mu)
  X^{l_zl_1}_{lm'}(\hat{\bm{z}},\hat{\bm{k}})
  \label{eq:24}
\end{multline}
for the first-order propagator, and
\begin{multline}
  \hat{\Gamma}^{(2)\,l\,l_z;L}_{Xl_1l_2}(k_1,k_2;k,\mu)
  = (2l_z+1)(2l_1+1)(2l_2+1)\,g_{(l)}^{mm'}
  \\ \times
  \int \frac{d^2\hat{z}}{4\pi} \frac{d^2\hat{k}_1}{4\pi}
  \frac{d^2\hat{k}_2}{4\pi}\,
  \hat{\Gamma}^{(2)}_{Xlm}(\bm{k}_1,\bm{k}_2;\hat{\bm{z}};k,\mu)
  X^{l_zl_1l_2}_{L;lm'}(\hat{\bm{z}},\hat{\bm{k}}_1,\hat{\bm{k}}_2)
  \label{eq:25}
\end{multline}
for the second-order propagator. In the above equations, the angular
integrations on the rhs in variables $k$ and $\mu$ are formally fixed,
as if they do not depend on $\hat{\bm{k}}$, $\hat{\bm{k}}_1$,
$\hat{\bm{k}}_2$ and $\hat{\bm{z}}$. The expressions of
Eqs.~(\ref{eq:24}) and (\ref{eq:25}) should be considered as
formal, and should only be used in order to invert the expansions of
Eqs.~(\ref{eq:18}) and (\ref{eq:19}), formally fixing the
variables $k$ and $\mu$ on both sides of the equations.

In practice, one can always represent the propagators with polypolar
spherical harmonics in the form of Eqs.~(\ref{eq:18}) and
(\ref{eq:19}), and can readily read the expression of the invariant
functions from the results. The invariant propagators in real space,
Eqs.~(\ref{eq:16}) and (\ref{eq:17}), correspond to
$\hat{\Gamma}^{(1)}_{Xl}(k) = \hat{\Gamma}^{(1)\,l0}_{Xl}(k)$,
$\hat{\Gamma}^{(2)\,l}_{Xl_1l_2}(k_1,k_2) =
\hat{\Gamma}^{(2)\,l\,0;l}_{Xl_1l_2}(k_1,k_2)$ when the propagators do
not contain redshift-space distortions.

\subsection{\label{subsec:FirstOrder}
  First-order propagators with loop corrections
}

\subsubsection{\label{subsubsec:CartProp}
  The propagators of integrated perturbation theory
}

The first-order and second-order propagators,
$\hat{\Gamma}^{(1)}_{Xlm}(\bm{k})$ and
$\hat{\Gamma}^{(2)}_{Xlm}(\bm{k}_1,\bm{k}_2)$ are evaluated by the
iPT. The results are given by (Sec.~III~A of Paper~I)
\begin{multline}
  \hat{\Gamma}_{Xlm}^{(1)}(\bm{k})
  = c_{Xlm}^{(1)}(\bm{k}) +
  \left[\bm{k}\cdot\bm{L}_1(\bm{k})\right] c_{Xlm}^{(0)}
  \\
  + \int\frac{d^3p}{(2\pi)^3} P_\mathrm{L}(p)
  \biggl\{
    \left[\bm{k}\cdot\bm{L}_1(-\bm{p})\right]
    c_{Xlm}^{(2)}(\bm{k},\bm{p})
    \\
      + \left[\bm{k}\cdot\bm{L}_1(-\bm{p})\right]
      \left[\bm{k}\cdot\bm{L}_1(\bm{k})\right]
      c_{Xlm}^{(1)}(\bm{p})
\\
      + \left[\bm{k}\cdot\bm{L}_2(\bm{k},-\bm{p})\right]
      c_{Xlm}^{(1)}(\bm{p})
      \\
      + \frac{1}{2}
      \left[\bm{k}\cdot\bm{L}_3(\bm{k},\bm{p},-\bm{p})\right]
      c_{Xlm}^{(0)}
      \\
      + \left[\bm{k}\cdot\bm{L}_1(\bm{p})\right]
      \left[\bm{k}\cdot\bm{L}_2(\bm{k},-\bm{p})\right]
       c_{Xlm}^{(0)}
  \biggr\}
\label{eq:26}
\end{multline}
for the first-order propagator, where $P_\mathrm{L}(k)$ is the linear
power spectrum of the mass density field, and
\begin{multline}
  \hat{\Gamma}^{(2)}_{Xlm}(\bm{k}_1,\bm{k}_2) = 
  c^{(2)}_{Xlm}(\bm{k}_1,\bm{k}_2)
\\
  + \left[\bm{k}_{12}\cdot\bm{L}_1(\bm{k}_1)\right] c^{(1)}_{Xlm}(\bm{k}_2)
  + \left[\bm{k}_{12}\cdot\bm{L}_1(\bm{k}_2)\right] c^{(1)}_{Xlm}(\bm{k}_1)
\\
  +
  \left\{
    \left[\bm{k}_{12}\cdot\bm{L}_1(\bm{k}_1)\right]
    \left[\bm{k}_{12}\cdot\bm{L}_1(\bm{k}_2)\right]
    + \bm{k}_{12}\cdot\bm{L}_2(\bm{k}_1,\bm{k}_2)
  \right\} c^{(0)}_{Xlm}.
\label{eq:27}
\end{multline}
for the second-order propagator. It is sufficient to include one-loop
corrections only in the first-order propagator
$\hat{\Gamma}^{(1)}_{Xlm}$, and not in the second-order propagator
$\hat{\Gamma}^{(2)}_{Xlm}$, because the second-order propagator always
appears with loop integrals in evaluating the nonlinear power spectrum
\cite{Matsubara:2011ck,Matsubara:2013ofa}.

In the above equations, $c^{(0)}_X$, $c^{(1)}_{Xlm}(\bm{k})$, and
$c^{(2)}_{Xlm}(\bm{k}_1,\bm{k}_2)$ are the renormalized bias
functions, which are determined from complicated physics involving
nonlinear dynamics of galaxy formation, etc., and the definitions of
these functions are given in Sec.~III~A of Paper~I. The vector
functions $\bm{L}_n(\bm{k}_1,\ldots,\bm{k}_n)$ are the kernel
functions of the Lagrangian perturbation theory. In real space, they
are explicitly given by \cite{Catelan:1994ze,Catelan:1996hw}
\begin{align}
  &\bm{L}_1(\bm{k})
  = \frac{\bm{k}}{k^2},
\label{eq:28}\\
  &\bm{L}_2(\bm{k}_1,\bm{k}_2)
  = \frac37 \frac{\bm{k}_{12}}{{k_{12}}^2}
  \left[1 - \left(\frac{\bm{k}_1 \cdot \bm{k}_2}{k_1 k_2}\right)^2\right],
\label{eq:29}\\
&  \bm{L}_3(\bm{k}_1,\bm{k}_2,\bm{k}_3) =
  \frac13
  \left[\tilde{\bm{L}}_3(\bm{k}_1,\bm{k}_2,\bm{k}_3) + \mathrm{cyc.}\right],
\label{eq:30}\\
& \tilde{\bm{L}}_3(\bm{k}_1,\bm{k}_2,\bm{k}_3)
\nonumber\\
& \quad
  = \frac{\bm{k}_{123}}{{k_{123}}^2}
  \left\{
      \frac57
      \left[1 - \left(\frac{\bm{k}_1 \cdot \bm{k}_2}{k_1 k_2}\right)^2\right]
      \left[1 - \left(\frac{\bm{k}_{12} \cdot \bm{k}_3}
          {{k}_{12} k_3}\right)^2\right]
  \right.
\nonumber\\
& \qquad\quad
  \left.
  - \frac13
  \left[
      1 - 3\left(\frac{\bm{k}_1 \cdot \bm{k}_2}{k_1 k_2}\right)^2
      +\, 2 \frac{(\bm{k}_1 \cdot \bm{k}_2)(\bm{k}_2 \cdot \bm{k}_3)
        (\bm{k}_3 \cdot \bm{k}_1)}{{k_1}^2 {k_2}^2 {k_3}^2}
  \right]\right\}
\nonumber\\
  & \qquad
  + \frac{3}{7}\frac{\bm{k}_{123}}{{k_{123}}^2}
    \times
    \frac{(\bm{k}_1\times\bm{k}_{23})(\bm{k}_1\cdot\bm{k}_{23})}{{k_1}^2{k_{23}}^2}
    \left[1 - \left(\frac{\bm{k}_2 \cdot \bm{k}_3}{k_2 k_3}\right)^2\right],
\label{eq:31}
\end{align}
where $\bm{k}_{12}=\bm{k}_1+\bm{k}_2$,
$\bm{k}_{123}=\bm{k}_1+\bm{k}_2+\bm{k}_3$, and $+\,\mathrm{cyc.}$
corresponds to the two terms which are added with cyclic permutations
of each previous term. Weak dependencies on the time in the kernels
are neglected \cite{Bernardeau:2001qr,Matsubara:2015ipa}. In
Ref.~\cite{Matsubara:2015ipa}, complete expressions of the
displacement kernels of Lagrangian perturbation theory up to the
seventh order including the transverse parts are explicitly given,
together with a general way of recursively deriving the kernels
including weak dependencies on the time in general cosmology and
subleading growing modes. The redshift-space distortions can be simply
taken into account as well in the Lagrangian perturbation theory, just
replacing the displacement kernel in real space given above with the
linearly mapped kernels \cite{Matsubara:2007wj}
\begin{equation}
  \bm{L}_n \rightarrow \bm{L}^\mathrm{s}_n = \bm{L}_n +
  nf\left(\hat{\bm{z}}\cdot\bm{L}_n\right)\hat{\bm{z}},
\label{eq:32}
\end{equation}
where $f=d\ln D/d\ln a = \dot{D}/HD$ is the linear growth rate, $D(t)$
is the linear growth factor, $a(t)$ is the scale factor, and
$H(t) = \dot{a}/a$ is the time-dependent Hubble parameter, and the
unit vector $\hat{\bm{z}}$ denotes the line-of-sight direction, as
already mentioned above.

The first-order propagators $\Gamma^{(1)}_{Xlm}$ in the lowest order
approximation, i.e., without loop corrections, are explicitly derived
in Sec.~IV~C~3 of Paper~I. The results are given by
\begin{equation}
  \hat{\Gamma}^{(1)}_{Xl}(k) =
  c^{(1)}_{Xl}(k) + \delta_{l0} c^{(0)}_X
  \label{eq:33}
\end{equation}
in real space, and
\begin{equation}
  \hat{\Gamma}^{(1)l\,l_z}_{Xl_1}(k,\mu) =
  \delta_{l_z0} \delta_{l_1l}
  \left[
    c^{(1)}_{Xl}(k) + \delta_{l0} (1+f\mu^2) c^{(0)}_X
  \right].
  \label{eq:34}
\end{equation}
in redshift space. In the last expression in redshift space, the
dependence on the direction cosine $\mu$ is kept unexpanded into
spherical harmonics. Another expression with the complete expansion is
given in Sec.~IV~C~1 and 2 of Paper~I. Below we consider the
generalization of the lowest-order results, and derive necessary
propagators for evaluating one-loop corrections in the nonlinear power
spectrum.

\subsubsection{\label{subsubsec:LoopPropReal}
  Real space
}

In the expression of propagators,
Eqs.~(\ref{eq:26})--(\ref{eq:31}), many scalar products
between pairs of wave vectors appear, and they can always be
represented by the spherical harmonics by applying the addition
theorem of the spherical harmonics,
\begin{equation}
  P_l(\bm{n}\cdot\bm{n}')
  = C_{lm}^*(\bm{n}) C_{lm}(\bm{n}'),
  \label{eq:35}
\end{equation}
where $\bm{n}$, $\bm{n}'$ are normal vectors, and $P_l(x)$ is the
Legendre polynomial. Einstein's summation convention for the index $m$
is applied as before. For example, we have
\begin{equation}
  \bm{k}\cdot\bm{L}_1(\pm \bm{p})
  = \pm \frac{\bm{k}\cdot\bm{p}}{p^2}
  = \pm \frac{k}{p}
  C_{1m}^*(\hat{\bm{k}}) C_{1m}(\hat{\bm{p}}).
  \label{eq:36}
\end{equation}
Similarly, all the directional dependencies on the wave vectors are
expanded into the spherical harmonics. In such reductions,
representing simple polynomials by Legendre polynomials, such as
\begin{align}
  &
    1 = P_0(x), \ \ 
    x = P_1(x), \ \ 
    x^2 = \frac{1}{3} P_0(x) + \frac{2}{3} P_2(x),
    \nonumber\\
  &
    x^3 = \frac{3}{5} P_1(x) + \frac{2}{5} P_3(x), \ \ 
    x^4 = \frac{1}{5} P_0(x) + \frac{4}{7} P_2(x) + \frac{8}{35} P_4(x),
  \label{eq:37}
\end{align}
are useful.

A nonpolynomial factor $|\bm{k}\pm\bm{p}|^{-2}$ appears in the
one-loop integrations in Eq.~(\ref{eq:26}) through the factors
\begin{align}
  \bm{k}\cdot\bm{L}_2(\bm{k},-\bm{p})
  &= \frac{3}{7} \frac{k^2 - \bm{k}\cdot\bm{p}}{|\bm{k}-\bm{p}|^2}
    \left[ 1 - (\hat{\bm{k}}\cdot\hat{\bm{p}})^2 \right],
  \label{eq:38}\\
  \bm{k}\cdot\bm{L}_3(\bm{k},\bm{p},-\bm{p})
  &= \frac{5}{21}
    \left(
    \frac{k^2}{|\bm{k}-\bm{p}|^2}
    + \frac{k^2}{|\bm{k}+\bm{p}|^2} 
    \right)
    \left[ 1 - (\hat{\bm{k}}\cdot\hat{\bm{p}})^2 \right]^2.
  \label{eq:39}
\end{align}
The directional dependence of this nonpolynomial factor can also be
expanded by the spherical harmonics, as we have a formula,
\begin{multline}
  \frac{1}{|\bm{k}\mp\bm{p}|^2} =
  \sum_{l=0}^\infty (\pm 1)^l(2l+1)
  C_{lm}^*(\hat{\bm{k}}) C_{lm}(\hat{\bm{p}})
  \\ \times
  \int_0^\infty \!r\,dr\,j_l(kr) j_l(pr),
  \label{eq:40}
\end{multline}
where $j_l(x)$ is the spherical Bessel function. The above equation
can be derived by simply rewriting the lhs as
\begin{align}
  \frac{1}{|\bm{k}\mp\bm{p}|^2}
  &=
    \int d^3q\,
    \frac{1}{q^2}
    \delta_\mathrm{D}^3(\bm{k}\mp\bm{p}-\bm{q})
    \nonumber\\
  &=
    \int d^3r\,e^{-i(\bm{k}\mp\bm{p})\cdot\bm{r}}
    \int\frac{d^3q}{(2\pi)^3}\,\frac{e^{i\bm{q}\cdot\bm{r}}}{q^2}.
  \label{eq:41}
\end{align}
The last integral over $\bm{q}$ is the Green's function of the
Laplacian and equals to $(4\pi r)^{-1}$, and the exponential factor
is expanded into plane-wave expansion,
\begin{equation}
  e^{\pm i\bm{k}\cdot\bm{r}} =
  \sum_{l=0}^\infty (\pm i)^l (2l+1) j_l(kr)
  C_{lm}^*(\hat{\bm{k}}) C_{lm'}(\hat{\bm{r}}).
  \label{eq:42}
\end{equation}
Integrating over the angular part of $\bm{r}$, and applying the
orthonormality relation of spherical harmonics, Eq.~(\ref{eq:13}),
it is straightforward to derive Eq.~(\ref{eq:40}).

Substituting Eq.~(\ref{eq:40}) into Eqs.~(\ref{eq:38}) and
(\ref{eq:39}), representing the polynomials of scalar products by
Legendre polynomials, and applying the addition theorem,
Eq.~(\ref{eq:35}), all the angular dependencies in
Eqs.~(\ref{eq:38}) and (\ref{eq:39}) on wave vectors $\bm{k}$ and
$\bm{p}$ are represented by spherical harmonics,
$C_{lm}(\hat{\bm{k}})$ and $C_{lm}(\hat{\bm{p}})$. The angular
dependencies of renormalized bias functions are also represented by
spherical harmonics just in a similar way to Eqs.~(\ref{eq:1}) and
(\ref{eq:2}),
\begin{align}
  c^{(1)}_{Xlm}(\bm{k})
  &= \frac{(-1)^l}{\sqrt{2l+1}}\,
  c^{(1)}_{Xl}(k) C_{lm}(\hat{\bm{k}}),
  \label{eq:43}\\
  c^{(2)}_{Xlm}(\bm{k}_1,\bm{k}_2)
  &= \sum_{l_1,l_2}
  c^{(2)\,l}_{Xl_1l_2}(k_1,k_2)
  X^{l_1l_2}_{lm}(\hat{\bm{k}}_1,\hat{\bm{k}}_2).
  \label{eq:44}
\end{align}
Thus the angular integration over $\bm{p}$ in the loop integral of
Eq.~(\ref{eq:26}) is analytically evaluated, where a formula of
the product of spherical harmonics,
\begin{multline}
  C_{l_1m_1}(\bm{n}) C_{l_2m_2}(\bm{n})
  =
  \sum_{l} (2l+1)
  \begin{pmatrix}
    l_1 & l_2 & l \\
    0 & 0 & 0
  \end{pmatrix}
  \\ \times
  \left(l_1\,l_2\,l\right)_{m_1m_2}^{\phantom{m_1m_2}m}
  C_{lm}(\bm{n}),
  \label{eq:45}
\end{multline}
is employed when necessary. The $3j$-symbol with vanishing azimuthal
indices $(l_1\,l_2\,l_3)_{000}$ is nonzero only when
$l_1+l_2+l_3=\mathrm{even}$. Thus we have an identity,
\begin{equation}
  (-1)^{l_1+l_2+l_3}
  \begin{pmatrix}
    l_1 & l_2 & l_3 \\
    0 & 0 & 0
  \end{pmatrix}
  =
  \begin{pmatrix}
    l_1 & l_2 & l_3 \\
    0 & 0 & 0
  \end{pmatrix},
  \label{eq:46}
\end{equation}
which we frequently use in the following calculations. 

After straightforward calculations described above, all the angular
integrations over $\hat{\bm{p}}$ in the loop integral of
Eq.~(\ref{eq:26}) are analytically performed. After lengthy but
straightforward calculations, the corresponding result of the
invariant propagator is obtained, and the result is given by
\begin{widetext}
\begin{multline}
  \hat{\Gamma}^{(1)}_{Xl}(k) =
  c^{(1)}_{Xl}(k) + \delta_{l0} c^{(0)}_X
  - \frac{\delta_{l1}}{3} Q^{(1)}_1(k)
  - \frac{(-1)^l}{3} \sqrt{2l+1}
  \sum_{l'}
  \begin{pmatrix}
    1 & l & l' \\
    0 & 0 & 0
  \end{pmatrix}
  Q^{(2)}_{ll'}(k)
  + \frac{1}{7} \delta_{l0} c^{(0)}_X
  \left[
    \frac{5}{3} R_1(k) + 3 R_2(k)
  \right]
  \\
  + \frac{2}{7}
  \sum_{l'}  (2l'+1)
  \left\{
    \left[
      \frac{\delta_{l'l}}{2l+1} -
      \begin{pmatrix}
        2 & l & l' \\
        0 & 0 & 0
      \end{pmatrix}^2
    \right]
    R^{(0)}_{ll'}(k)
      + \frac{3}{5}
    \left[
      \begin{pmatrix}
        3 & l & l' \\
        0 & 0 & 0
      \end{pmatrix}^2
      -
      \begin{pmatrix}
        1 & l & l' \\
        0 & 0 & 0
      \end{pmatrix}^2
    \right]
    R^{(1)}_{ll'}(k)
  \right\},
  \label{eq:47}
\end{multline}
where
\begin{align}
  Q^{(1)}_l(k)
  &\equiv
    \int \frac{p^2dp}{2\pi^2} P_\mathrm{L}(p) \frac{k}{p}
    c^{(1)}_{Xl}(p), \qquad
  Q^{(2)}_{ll'}(k) \equiv
    \int \frac{p^2dp}{2\pi^2} P_\mathrm{L}(p) \frac{k}{p}
    c^{(2)\,l}_{Xl'1}(k,p),
    \label{eq:48}\\
  R^{(n)}_{ll'}(k)
  &\equiv k^2 \int r\,dr\,j_{l'}(kr)
    \int \frac{p^2dp}{2\pi^2} P_\mathrm{L}(p)
    \left(\frac{p}{k}\right)^n j_{l'}(pr) c^{(1)}_{Xl}(p),
    \label{eq:49}\\
  R_1(k)
  &\equiv k^2 \int r\,dr\,
    \left[
    \frac{8}{15}j_0(kr)\xi^{(0)}_0(r)
    - \frac{16}{21}j_2(kr)\xi^{(0)}_2(r)
    + \frac{8}{35}j_4(kr)\xi^{(0)}_4(r)
    \right],
    \label{eq:50}\\
  R_2(k)
  &\equiv k^2 \int r\,dr\,
    \left\{
    - \frac{2}{15}j_0(kr)\xi^{(0)}_0(r)
    - \frac{2}{21}j_2(kr)\xi^{(0)}_2(r)
    + \frac{8}{35}j_4(kr)\xi^{(0)}_4(r)
    + \frac{2k}{5}
    \left[
    j_1(kr)\xi^{(-1)}_1(r) - j_3(kr)\xi^{(-1)}_3(r)
    \right]
    \right\},
    \label{eq:51}
\end{align}
\end{widetext}
and
\begin{equation}
  \xi^{(n)}_l(r) \equiv
  \int \frac{k^2dk}{2\pi^2} P_\mathrm{L}(k)\,k^n j_l(kr).
  \label{eq:52}
\end{equation}
The integrals that appear in the above expressions are essentially
one-dimensional Hankel transforms, which can be numerically evaluated
with a one-dimensional fast Fourier transform (FFT) with a famous code
\textsc{FFTLog} developed by Hamilton \cite{Hamilton:1999uv}. In the
scalar case $l=0$ without bias, only the last term of the first line
in Eq.~(\ref{eq:47}) is present in one-loop corrections, and the
relevant functions $R_1(k)$ and $R_2(k)$ represented in the form of
Hankel transforms are essentially the same as those previously derived
in the FFT-PT formalism
\cite{Schmittfull:2016jsw,Schmittfull:2016yqx}.%
\footnote{The normalization of the functions $R_1(k)$ and $R_2(k)$ are
  different from those defined in Ref.~\cite{Schmittfull:2016jsw} by
  an extra factor $P_\mathrm{L}(k)$.} Thus, the method to calculate
loop corrections in the present formalism can be seen as a natural
generalization of the FFT-PT formalism (or its variant, FAST-PT
formalism \cite{McEwen:2016fjn,Fang:2016wcf}) to the case of
tensor-valued biased fields.

\subsubsection{\label{subsubsec:LoopPropRed}
  Redshift space
}

The first-order propagator in redshift space in the lowest-order
approximation is given by Eq.~(\ref{eq:34}). In redshift space,
however, the angular dependence of the total wave vector
$\bm{k}=\bm{k}_1+\cdots +\bm{k}_n$ with respect to the line of sight
$\hat{\bm{z}}$ does not have to be expanded with the spherical
harmonics in evaluating the power spectrum and higher-order
polyspectra, because the dependence is always factored out in the
expressions as we will explicitly see below. Thus, we keep the
directional cosine $\mu = \hat{\bm{z}}\cdot\hat{\bm{k}}$ of the total
wave vector in the expression, without expanding into spherical
tensors. For example, the angular dependence in the second term on the
rhs of Eq.~(\ref{eq:34}) is not expanded into spherical harmonics.
Although it is always possible to expand the directional dependencies
in $k$ and $\mu$ into spherical harmonics, the expressions become much
more cumbersome. Below we keep the variables $k$ and $\mu$ unexpanded
as much as possible.

The calculation of the first-order propagator with one-loop
corrections in redshift space is also straightforward, with almost the
same techniques employed above in the calculations in real space. One
can just replace the perturbation kernels in Eq.~(\ref{eq:26})
with Eq.~(\ref{eq:32}). As naturally expected, the resulting
expressions become lengthier. After some lengthy but straightforward
calculations, the result is given by
\begin{widetext}
  \begin{multline}
    \hat{\Gamma}^{(1)\,l\,l_z}_{Xl_1}(k,\mu) =
    \delta_{l_z0} \delta_{l_1l}
    \left[
    c^{(1)}_{Xl}(k) +
    \delta_{l0} (1+f\mu^2) c^{(0)}_X
    \right]
    - \frac{\delta_{l1}}{3} (1+f\mu^2)
    \left(
      \delta_{l_z0} \delta_{l_11}
      + f\mu\,\delta_{l_z1}\delta_{l_10}
    \right)
    Q^{(1)}_1(k)
    \\
    - \frac{(-1)^l}{3} \sqrt{2l+1}
      \delta_{l_z0} \delta_{l_1l}
      \sum_{l'}
      \begin{pmatrix}
        1 & l & l' \\ 0 & 0 & 0
      \end{pmatrix}
      Q^{(2)}_{ll'}(k)
      + \frac{1}{3} f\mu (-1)^{l+l_1} \delta_{l_z1}
      Q^{(2)}_{ll_1}(k)
      \\
      + \frac{1}{7} \delta_{l0} \delta_{l_z0} \delta_{l_10} c^{(0)}_X
      \left\{
      \left[
      \frac{5}{3} + (5-3f)f\mu^2 + 3 f^2 \mu^4
      \right] R_1(k)
      + 3(1+f\mu^2)(1+2f\mu^2) R_2(k)
      \right\}
    \\
    +
    \frac{2}{7}
    \delta_{l_z0} \delta_{l_1l}
    \sum_{l'}  (2l'+1)
    \left\{
      (1+2f\mu^2)
      \left[
        \frac{\delta_{l'l}}{2l+1} -
        \begin{pmatrix}
          2 & l & l' \\
          0 & 0 & 0
        \end{pmatrix}^2
      \right]
      R^{(0)}_{ll'}(k)
      + \frac{3}{5}
      \left[
        \begin{pmatrix}
          3 & l & l' \\
          0 & 0 & 0
        \end{pmatrix}^2
        -
        \begin{pmatrix}
          1 & l & l' \\
          0 & 0 & 0
        \end{pmatrix}^2
      \right]
      R^{(1)}_{ll'}(k)
    \right\}
    \\
    - \frac{4}{7} f\mu\,\delta_{l_z1}
      (-1)^l \frac{2l_1+1}{\sqrt{2l+1}}
      \begin{pmatrix}
        1 & l & l_1 \\ 0 & 0 & 0
      \end{pmatrix}
      \sum_{l'} (2l'+1)
      \left[
        \frac{\delta_{l'l_1}}{2l_1+1}
        - 
        \begin{pmatrix}
          2 & l_1 & l' \\ 0 & 0 & 0
        \end{pmatrix}^2
      \right]
      R^{(1)}_{ll'}(k).
    \label{eq:53}
  \end{multline}
\end{widetext}
Substituting $l_1=l$, $l_z=0$ and $f=0$, the above result for the
first-order propagator in redshift space reduces to that in real
space, Eq.~(\ref{eq:47}):
$\hat{\Gamma}^{(1)}_{Xl}(k) =
\left.\hat{\Gamma}^{(1)\,l\,0}_{Xl}(k,\mu)\right|_{f=0}$.

\subsection{\label{subsec:SecondOrder}
  Second-order propagators
}

To evaluate the one-loop corrections in the power spectrum, it is not
necessary to evaluate the loop corrections to second-order
propagators, because the second-order propagators appear only with
second-order terms in the linear power spectrum (see
Refs.~\cite{Matsubara:2011ck,Matsubara:2013ofa} for a relevant
diagram). The calculations of the second-order propagators are
straightforward as in the first-order case. The propagator in real
space is given by Eq.~(\ref{eq:27}), and that in redshift space is
given by the same equation with substitutions of
Eq.~(\ref{eq:32}). The procedure to obtain analytic expressions of
the invariant propagators of the first order employed above can be
almost equally applied to obtain the expressions of the invariant
propagators of second order, and the calculations are fairly
straightforward even if they are lengthy. We thus just summarize the
results below.

In real space, the invariant propagator of the second order is given by
\begin{widetext}
  \begin{multline}
    \hat{\Gamma}^{(2)\,l}_{Xl_1l_2}(k_1,k_2)
    = c^{(2)\,l}_{Xl_1l_2}(k_1,k_2)
    +
    \left[
      \delta_{l_1l} \delta_{l_20}
      + (-1)^l\delta_{l_21} \frac{k_1}{k_2}
       \frac{2l_1+1}{\sqrt{2l+1}}
      \begin{pmatrix}
        1 & l & l_1 \\ 0 & 0 & 0
      \end{pmatrix}
    \right] c^{(1)}_{Xl}(k_1) + (1\leftrightarrow 2)
    \\
    + \delta_{l0}
    \left[
      \frac{34}{21} \delta_{l_10} \delta_{l_20}
      + \frac{8\sqrt{5}}{21} \delta_{l_12} \delta_{l_22}
      - \sqrt{3} 
      \left(\frac{k_2}{k_1} + \frac{k_1}{k_2}\right)
      \delta_{l_11} \delta_{l_21}
     \right] c^{(0)}_X,
    \label{eq:54}
  \end{multline}
where $+\,(1\leftrightarrow 2)$ represents a term with interchanged
subscripts of the previous term. The above result apparently satisfies
interchange symmetry of Eq.~(\ref{eq:15}).

In redshift space, the invariant propagator of second order is given
by
  \begin{multline}
    \hat{\Gamma}^{(2)\,l\,l_z;L}_{Xl_1l_2}(k_1,k_2;k,\mu)
    = \delta_{Ll} \delta_{l_z0}c^{(2)\,l}_{Xl_1l_2}(k_1,k_2)
    \\
    +
    \left\{
      \delta_{Ll} \delta_{l_z0}
      \left[
        \delta_{l_1l} \delta_{l_20}
        + (-1)^l \delta_{l_21} \frac{k_1}{k_2} \frac{2l_1+1}{\sqrt{2l+1}}
        \begin{pmatrix}
          1 & l & l_1 \\ 0 & 0 & 0
        \end{pmatrix}
      \right]
      + (-1)^{l+L}
      \delta_{l_z1} \delta_{l_1l} \delta_{l_21} \delta^\triangle_{1lL}
      \frac{f \mu k}{k_2}
      \sqrt{\frac{2L+1}{2l+1}}
    \right\} c^{(1)}_{Xl}(k_1)
    \\
    +
    \left\{
      \delta_{Ll} \delta_{l_z0}
      \left[
        \delta_{l_10} \delta_{l_2l}
        + (-1)^l \delta_{l_11} \frac{k_2}{k_1} \frac{2l_2+1}{\sqrt{2l+1}}
        \begin{pmatrix}
          1 & l & l_2 \\ 0 & 0 & 0
        \end{pmatrix}
      \right]
      - \delta_{l_z1} \delta_{l_11} \delta_{l_2l} \delta^\triangle_{1lL}
      \frac{f \mu k}{k_1}
      \sqrt{\frac{2L+1}{2l+1}}
    \right\} c^{(1)}_{Xl}(k_2)
    \\
    + \delta_{l0}
    \left\{
      \delta_{L0} \delta_{l_z0}
      \left[
        \left(
          \frac{34}{21} + \frac{4}{7} f \mu^2
        \right)
        \delta_{l_10} \delta_{l_20}
        + \frac{8\sqrt{5}}{21}
        \left(1 - \frac{3}{2} f\mu^2\right)
        \delta_{l_12} \delta_{l_22}
        - \sqrt{3} 
        \left(\frac{k_2}{k_1} + \frac{k_1}{k_2}\right)
        \delta_{l_11} \delta_{l_21}
      \right]
      \right.
      \\
      \left.
      - \frac{f\mu k}{k_1} \delta_{L1} \delta_{l_z1} \delta_{l_11}
      \left[
        \sqrt{3}\,\delta_{l_20} - (2l_2+1)
        \begin{pmatrix}
          1 & 1 & l_2 \\ 0 & 0 & 0
        \end{pmatrix}
      \right] + (1\leftrightarrow 2)
      + \frac{f^2 \mu^2 k^2}{k_1k_2}
      \delta_{Ll_z} \delta_{l_11} \delta_{l_21}
      (2l_z+1)
      \begin{pmatrix}
        1 & 1 & l_z \\ 0 & 0 & 0
      \end{pmatrix}
     \right\} c^{(0)}_X.
    \label{eq:55}
  \end{multline}
\end{widetext}
This result apparently satisfies the interchange symmetry of
Eq.~(\ref{eq:21}). Substituting $l_z=0$, $L=l$ and $f=0$, the above
result for the second-order propagator in redshift space,
Eq.~(\ref{eq:55}), reduces to that in real space,
Eq.~(\ref{eq:54}):
$\hat{\Gamma}^{(2)\,l}_{Xl_1l_2}(k_1,k_2) =
\left.\hat{\Gamma}^{(2)\,l\,0;l}_{Xl_1l_2}(k_1,k_2;k,\mu)\right|_{f=0}$.

\section{\label{sec:OneLoopPS}
  Loop corrections to the nonlinear power spectrum of tensor fields
}

\subsection{\label{subsec:PowerSpec}
  The power spectrum of tensor fields
}

In Sec.~III~A of Paper~I, the definition of cosmological tensor fields
$F_{Xi_1i_2\cdots i_l}(\bm{x})$ is given, where the index $X$
specifies the kind of objects such as galaxies we observe, and
$i_1,\ldots,i_l$ are Cartesian indices of the rank-$l$ tensor which is
attributed to the objects. The tensor fields are decomposed into
irreducible tensors $F_{Xlm}(\bm{x})$ on the spherical basis in
general. The power spectrum $P^{(l_1l_2)}_{X_1X_2m_1m_2}(\bm{k})$ of
irreducible tensor fields is defined by (Sec.~V~A of Paper~I)
\begin{multline}
  \left\langle
    F_{X_1l_1m_1}(\bm{k}_1)
    F_{X_2l_2m_2}(\bm{k}_2)
  \right\rangle_\mathrm{c}
  \\
  =
  (2\pi)^3\delta_\mathrm{D}^3(\bm{k}_1+\bm{k}_2)
  P^{(l_1l_2)}_{X_1X_2m_1m_2}(\bm{k}_1),
  \label{eq:56}
\end{multline}
where $\langle\cdots\rangle_\mathrm{c}$ indicates the connected part
of the two-point function, and the appearance of the delta function is
due to translational symmetry. The above power spectrum depends on the
coordinates system, and represented by rotationally invariant power
spectra. Below we first summarize the essential equations for the
power spectrum and correlation function, which are mainly derived in
Sec.~V~A of Paper~I. Readers who have already read the corresponding
section of Paper~I can skip this subsection and jump directly to
Sec.~\ref{subsec:OneLoopPS}.

\subsubsection{Real space}

In real space, the rotational symmetry requires that
the power spectrum should have a particular form,
\begin{equation}
  P^{(l_1l_2)}_{X_1X_2m_1m_2}(\bm{k}) =
  i^{l_1+l_2}
  \sum_l \left(l_1\,l_2\,l\right)_{m_1m_2}^{\phantom{m_1m_2}m}
  C_{lm}(\hat{\bm{k}}) P^{l_1l_2;l}_{X_1X_2}(k).
    \label{eq:57}
\end{equation}
and the function $P^{l_1l_2;l}_{X_1X_2}(k,\mu)$ is the invariant power
spectrum. The invariant power spectrum defined above is shown to be a
real function. Noting an orthonormality relation of the spherical
harmonics, Eq.~(\ref{eq:13}) and that of $3j$-symbol,
\begin{equation}
    \left(l_1\,l_2\,l\right)_{m_1m_2}^{\phantom{m_1m_2}m}
    \left(l_1\,l_2,l'\right)^{m_1m_2}_{\phantom{m_1m_2}m'}
    =
    \frac{(-1)^{l+l_1+l_2} \delta_{ll'}}{2l+1}
    \delta^m_{m'} \delta^\triangle_{l\,l_1l_2},
  \label{eq:58}
\end{equation}
the inverse relation of Eq.~(\ref{eq:57}) is given by
\begin{multline}
  P^{l_1l_2;l}_{X_1X_2}(k) =
    i^{l_1+l_2}(-1)^l    
  (2l+1)
  \\ \times
  \left(l_1\,l_2\,l\right)^{m_1m_2}_{\phantom{m_1m_2}m}
  \int \frac{d^2\hat{k}}{4\pi}\, P^{(l_1l_2)}_{X_1X_2m_1m_2}(\bm{k})
  C_{lm}^*(\hat{\bm{k}}). 
  \label{eq:59}
\end{multline}

The power spectrum in real space has an interchange symmetry,
\begin{equation}
  P^{(l_2l_1)}_{X_2X_1m_2m_1}(\bm{k})
  = P^{(l_1l_2)}_{X_1X_2m_1m_2}(-\bm{k}),
  \label{eq:60}
\end{equation}
and the corresponding symmetry for the invariant spectrum is given by
\begin{equation}
  P^{l_2l_1;l}_{X_2X_1}(k) =
  (-1)^{l_1+l_2} P^{l_1l_2;l}_{X_1X_2}(k).
  \label{eq:61}
\end{equation}

\subsubsection{Redshift space}

In redshift space, the rotational symmetry requires that the power
spectrum should have a particular form,
\begin{multline}
  P^{(l_1l_2)}_{X_1X_2m_1m_2}(\bm{k};\hat{\bm{z}}) =
    i^{l_1+l_2}
  \sum_{l,l_z,L}
  (-1)^L \sqrt{\{L\}}\,
  \left(l_1\,l_2\,L\right)_{m_1m_2}^{\phantom{m_1m_2}M} \\ \times
  X^{l_zl}_{LM}(\hat{\bm{z}},\hat{\bm{k}})
  P^{l_1l_2;l\,l_z;L}_{X_1X_2}(k,\mu),
  \label{eq:62}
\end{multline}
due to rotational symmetry, and the function
$P^{l_1l_2;l}_{X_1X_2}(k,\mu)$ is the invariant power spectrum. If the
directional dependencies are completely expanded into spherical
harmonics, the invariant spectrum does not depend on the directional
cosine, $\mu=\hat{\bm{z}}\cdot\hat{\bm{k}}$, while it is frequently
convenient to explicitly keep some part of the dependence on $\mu$
unexpanded into spherical harmonics, just as in the case of the
invariant propagator in redshift space.

On one hand, when the invariant spectrum does not depend on the
direction cosine $\mu$, one can uniquely obtain the inverse relation
of the above equation. Using orthonormality relations of bipolar
spherical harmonics and the $3j$-symbol, Eqs.~(\ref{eq:14}) and
(\ref{eq:58}), we derive
\begin{multline}
  P^{l_1l_2;l\,l_z;L}_{X_1X_2}(k) =
  i^{l_1+l_2}   
  (2l+1)(2l_z+1) \sqrt{(2L+1)}
  \\ \times
  \left(l_1\,l_2\,L\right)^{m_1m_2M}
  \int \frac{d^2\hat{z}}{4\pi} \frac{d^2\hat{k}}{4\pi}\,
  P^{(l_1l_2)}_{X_1X_2m_1m_2}(\bm{k};\hat{\bm{z}})
  X^{l_zl}_{LM}(\hat{\bm{z}},\hat{\bm{k}}).
  \label{eq:63}
\end{multline}
On the other hand, when the invariant spectrum does depend on the
direction cosine, the expansion of Eq.~(\ref{eq:62}) is not unique, as
which part of the direction cosine is expanded into spherical
harmonics and which part is not can be arbitrarily chosen. However,
if one fixes which part, the expansion is formally inverted by a
similar equation of Eq.~(\ref{eq:63}), in which both the original
power spectrum $P^{(l_1l_2)}_{X_1X_2m_1m_2}$ and the invariant power
spectrum $P^{l_1l_2;l\,l_z;L}_{X_1X_2}$ explicitly depend on variables
$k$ and $\mu$ and the latter dependencies are formally fixed as if
they do not depend on the directions of $\hat{\bm{k}}$ and
$\hat{\bm{z}}$ in the integral on the lhs.

The power spectrum in redshift space has an interchange symmetry,
\begin{equation}
  P^{(l_2l_1)}_{X_2X_1m_2m_1}(\bm{k};\hat{\bm{z}}) 
  = P^{(l_1l_2)}_{X_1X_2m_1m_2}(-\bm{k};\hat{\bm{z}}),
  \label{eq:64}
\end{equation}
and the corresponding symmetry for the invariant spectrum is given by
\begin{equation}
  P^{l_2l_1;l\,l_z;L}_{X_2X_1}(k) =
  (-1)^{l_1+l_2+l+L} P^{l_1l_2;l\,l_z;L}_{X_1X_2}(k).
  \label{eq:65}
\end{equation}

\subsubsection{Correlation functions}

While the power spectrum is defined in Fourier space, the counterpart
in configuration space is the correlation function,
$\xi^{(l_1l_2)}_{X_1X_2m_1m_2}(\bm{x})$, which is defined by
\begin{equation}
  \left\langle
    F_{X_1l_1m_1}(\bm{x}_1)
    F_{X_2l_2m_2}(\bm{x}_2)
  \right\rangle_\mathrm{c} =
  \xi^{(l_1l_2)}_{X_1X_2m_1m_2}(\bm{x}_1-\bm{x}_2),
  \label{eq:66}
\end{equation}
where the tensor field $F_{Xlm}(\bm{x})$ on the lhs corresponds to a
variable in configuration space. On the rhs, the correlation function
is a function of the relative position vector between the two
positions. The correlation function and the power spectrum are related
by a three-dimensional Fourier transform as
\begin{equation}
  \xi^{(l_1l_2)}_{X_1X_2m_1m_2}(\bm{x}) =
  \int \frac{d^3k}{(2\pi)^3} e^{i\bm{k}\cdot\bm{x}}
  P^{(l_1l_2)}_{X_1X_2m_1m_2}(\bm{k}),
  \label{eq:67}
\end{equation}
which is generally known as the Wiener-Khinchin theorem.

First, we consider the correlation function in real space. Because of
the rotational symmetry, the correlation function should have a form,
\begin{equation}
  \xi^{(l_1l_2)}_{X_1X_2m_1m_2}(\bm{x}) =
    i^{l_1+l_2}
  \sum_l i^l \left(l_1\,l_2\,l\right)_{m_1m_2}^{\phantom{m_1m_2}m}
  C_{lm}(\hat{\bm{x}})\, \xi^{l_1l_2;l}_{X_1X_2}(x),
    \label{eq:68}
\end{equation}
and the last factor $\xi^{l_1l_2;l}_{X_1X_2}(r)$ corresponds to the
invariant correlation function. The factor $i^l$ is additionally
present compared to the corresponding Eq.~(\ref{eq:57}) of the power
spectrum, and the invariant correlation function with the above
definition is a real function. The invariant power spectra and
invariant correlation functions are related by a Hankel transform,
\begin{equation}
  \xi^{l_1l_2;l}_{X_1X_2}(x) =
  \int \frac{k^2dk}{2\pi^2} j_l(kx) 
  P^{l_1l_2;l}_{X_1X_2}(k),
  \label{eq:69}
\end{equation}
and its inverse relation,
\begin{equation}
  P^{l_1l_2;l}_{X_1X_2}(x) =
  4\pi \int r^2 dr j_l(kx) 
  \xi^{l_1l_2;l}_{X_1X_2}(x). 
  \label{eq:70}
\end{equation}

Similarly, the correlation function in redshift space is also
considered. The relation between the correlation function and the
power spectrum is just given by Eq.~(\ref{eq:67}) as well in redshift
space, and in this case both explicitly depend on the direction of the
line of sight, $\hat{\bm{z}}$:
\begin{equation}
  \xi^{(l_1l_2)}_{X_1X_2m_1m_2}(\bm{x};\hat{\bm{z}}) =
  \int \frac{d^3k}{(2\pi)^3} e^{i\bm{k}\cdot\bm{x}}
  P^{(l_1l_2)}_{X_1X_2m_1m_2}(\bm{k};\hat{\bm{z}}).
  \label{eq:71}
\end{equation}
Because of the rotational symmetry, we have
\begin{multline}
  \xi^{(l_1l_2)}_{X_1X_2m_1m_2}(\bm{x};\hat{\bm{z}}) =
    i^{l_1+l_2}
  \sum_{L} (-1)^L \sqrt{2L+1}
  \left(l_1\,l_2\,L\right)_{m_1m_2}^{\phantom{m_1m_2}M}
  \\ \times
  \sum_{l,l_z} i^l  X^{l_zl}_{LM}(\hat{\bm{z}},\hat{\bm{x}})
  \xi^{l_1l_2;l\,l_z;L}_{X_1X_2}(x),
  \label{eq:72}
\end{multline}
just as in the case of the power spectrum of Eq.~(\ref{eq:62}), and the
last factor $\xi^{l_1l_2;l\,l_z;L}_{X_1X_2}(r)$ corresponds to the
invariant correlation function.

When the invariant spectrum does not depend on the direction cosine
$\mu$, following the same procedure to derive Eqs.~(\ref{eq:69}) and
(\ref{eq:70}) in real space, we similarly have
\begin{align}
  \xi^{l_1l_2;l\,l_z;L}_{X_1X_2}(r)
  &=
  \int \frac{k^2dk}{2\pi^2} j_l(kr) 
    P^{l_1l_2;l\,l_z;L}_{X_1X_2}(k),
  \label{eq:73}\\
  P^{l_1l_2;l\,l_z;L}_{X_1X_2}(k)
  &=
  4\pi \int r^2 dr j_l(kr) 
  \xi^{l_1l_2;l\,l_z;L}_{X_1X_2}(r). 
  \label{eq:74}
\end{align}
However, when the invariant spectrum depends on the direction cosine
$\mu$, the above equations do not hold anymore. The treatment of the
latter case is described in Sec.~V~B of Paper~I.

\subsection{\label{subsec:OneLoopPS}
  The one-loop power spectra
}

Straightforwardly generalizing the original formalism of iPT
\cite{Matsubara:2011ck,Matsubara:2013ofa}, the nonlinear power
spectrum of tensor fields up to the one-loop approximation is given by
\begin{multline}
  P^{(l_1l_2)}_{X_1X_2m_1m_2}(\bm{k}) =
    i^{l_1+l_2}
  \Pi^2(\bm{k})
  \Biggl[
    \hat{\Gamma}^{(1)}_{X_1l_1m_1}(\bm{k})
    \hat{\Gamma}^{(1)}_{X_2l_2m_2}(-\bm{k})
    P_\mathrm{L}(k)
    \\
    + \frac{1}{2} \int \frac{d^3k'}{(2\pi)^3} \frac{d^3k''}{(2\pi)^3}
    (2\pi)^3\delta_\mathrm{D}^3(\bm{k}'+\bm{k}''-\bm{k})
    \\ \times
    \hat{\Gamma}^{(2)}_{X_1l_1m_1}(\bm{k}',\bm{k}'')
    \hat{\Gamma}^{(2)}_{X_2l_2m_2}(-\bm{k}',-\bm{k}'')
    P_\mathrm{L}(k') P_\mathrm{L}(k'')
  \Biggr],
  \label{eq:75}
\end{multline}
where the function $\Pi(\bm{k})$ is the resummation factor. Up to the
one-loop approximation, the function is explicitly given by
\begin{equation}
  \Pi(k) =
  \exp\left[
    -\frac{k^2}{12\pi^2}
    \int dp P_\mathrm{L}(p)
  \right],
  \label{eq:76}
\end{equation}
in real space, and 
\begin{equation}
  \Pi(k,\mu) =
  \exp\left\{
    -\frac{k^2}{12\pi^2}
    \left[ 1 + f(f+2)\mu^2 \right]
    \int dp P_\mathrm{L}(p)
  \right\},
  \label{eq:77}
\end{equation}
in redshift space, where $\mu \equiv \hat{\bm{z}}\cdot\hat{\bm{k}}$ is
the direction cosine between the wave vector and the line of sight.

In this paper, we assume the initial distributions of density
fluctuations are Gaussian, and primordial non-Gaussianity in the
initial condition is absent (the leading-order power spectrum in the
presence of primordial non-Gaussianity is already given in Paper I).
The first term in the square bracket on the rhs of Eq.~(\ref{eq:75})
is already given in Paper I. Denoting the corresponding invariant
spectrum of leading order as $P^{l_1l_2;l}_{X_1X_2\,[1]}(k)$ in real
space, the result is given by
\begin{multline}
  P^{l_1l_2;l}_{X_1X_2\,[1]}(k)
  = \frac{(-1)^{l_1} \{l\}}{\sqrt{(2l_1+1)(2l_2+1)}}
    \begin{pmatrix}
      l_1 & l_2 & l \\
      0 & 0 & 0
    \end{pmatrix}
    \\ \times
    \Pi^2(k)
    \hat{\Gamma}^{(1)}_{X_1l_1}(k)
    \hat{\Gamma}^{(1)}_{X_2l_2}(k)
    P_\mathrm{L}(k).
  \label{eq:78}
\end{multline}
In redshift space, the corresponding invariant spectrum is denoted as
$P^{l_1l_2;l\,l_z;L}_{X_1X_2\,[1]}(k,\mu)$ and is given by
\begin{multline}
  P^{l_1l_2;l\,l_z;L}_{X_1X_2\,[1]}(k,\mu) =
  (-1)^{l+l_z}
  \Pi^2(k,\mu) P_\mathrm{L}(k)
  \{l\} \{l_z\} \sqrt{\{L\}}
  \\ \times
  \sum_{l_{z1},l_{z2},l_1',l_2'}  (-1)^{l_2'}
  \begin{pmatrix}
    l_{z1} & l_{z2} & l_z \\
    0 & 0 & 0
  \end{pmatrix}
  \begin{pmatrix}
    l_1' & l_2' & l \\
    0 & 0 & 0
  \end{pmatrix}
  \begin{Bmatrix}
    l_{z1} & l_{z2} & l_z \\
    l_1' & l_2' & l \\
    l_1 & l_2 & L
  \end{Bmatrix}
  \\ \times
  \hat{\Gamma}^{(1)l_1l_{z1}}_{X_1l_1'}(k,\mu)
  \hat{\Gamma}^{(1)l_2l_{z2}}_{X_2l_2'}(k,-\mu),
  \label{eq:79}
\end{multline}
where the factor in front of the product of propagators is Wigner's
$9j$-symbol. The factors such as $2l+1$, $2l_1+1$, $2L+1$ etc.~quite
frequently appear throughout this paper, and we employ simplified
notations,
\begin{equation}
  \{l\} \equiv 2l+1,\ 
  \{l_1\} \equiv 2l_1+1,\ 
  \{L\} \equiv 2L+1,\ \mathrm{etc.}
  \label{eq:80}
\end{equation}
from here onward. Because of an identity for a special case of
$9j$-symbol \cite{Khersonskii:1988krb},
\begin{equation}
  \begin{Bmatrix}
    0 & 0 & 0 \\
    l_1' & l_2' & l_3' \\
    l_1 & l_2 & l_3
  \end{Bmatrix}
  = \frac{\delta_{l_1l_1'}\delta_{l_2l_2'}\delta_{l_3l_3'}}
  {\sqrt{\{l_1\}\{l_2\}\{l_3\}}},
  \label{eq:81}
\end{equation}
we readily see that the power spectrum in redshift space,
Eq.~(\ref{eq:79}), reduces to that in real space,
Eq.~(\ref{eq:78}), when we substitute $l_z = l_{z1} = l_{z2} = 0$.

\subsubsection{\label{subsubsec:PSreal}
  Real space
}

In order to evaluate the loop integral of the second term in the
square bracket on the rhs of Eq.~(\ref{eq:75}), we represent the
integrals with the delta function as
\begin{multline}
    \int \frac{d^3k'}{(2\pi)^3} \frac{d^3k''}{(2\pi)^3}
    (2\pi)^3\delta_\mathrm{D}^3(\bm{k}'+\bm{k}''-\bm{k})
    \times \cdots
    \\
    =
    \int d^3r\,e^{-i\bm{k}\cdot\bm{r}}
    \int \frac{d^3k'}{(2\pi)^3} \frac{d^3k''}{(2\pi)^3}
     e^{i(\bm{k}'+\bm{k}'')\cdot\bm{r}} \times \cdots,
  \label{eq:82}
\end{multline}
and apply plane-wave expansion of the exponential function,
Eq.~(\ref{eq:42}). Substituting Eq.~(\ref{eq:2}), the product of
second-order propagators in Eq.~(\ref{eq:75}) is given by a product
of invariant propagators and a product of bipolar spherical harmonics.
The product of bipolar spherical harmonics reduces to a single bipolar
spherical harmonics according to a formula
\cite{Khersonskii:1988krb,PaperI}
\begin{multline}
  X^{l_1'l_1''}_{l_1m_1}(\hat{\bm{k}}',\hat{\bm{k}}'')
  X^{l_2'l_2''}_{l_2m_2}(\hat{\bm{k}}',\hat{\bm{k}}'')
  =
  \sum_l (-1)^l \{l\}
  \left(l\,l_1\,l_2\right)^m_{\phantom{m}m_1m_2}
  \\ \times
  \sum_{l',l''} (-1)^{l'+l''} \{l'\}\{l''\}
  \begin{pmatrix}
    l_1' & l_2' & l' \\ 0 & 0 & 0
  \end{pmatrix}
  \begin{pmatrix}
    l_1'' & l_2'' & l'' \\ 0 & 0 & 0
  \end{pmatrix}
  \\ \times
  \begin{Bmatrix}
    l_1 & l_2 & l \\
    l_1' & l_2' & l' \\
    l_1'' & l_2'' & l''
  \end{Bmatrix}
  X^{l'l''}_{lm}(\hat{\bm{k}}',\hat{\bm{k}}'').
  \label{eq:83}
\end{multline}
For angular integrations over $\bm{k}'$ and $\bm{k}''$, we only need
the following equation:
\begin{multline}
  \int \frac{d^2\hat{k}'}{4\pi} \frac{d^2\hat{k}''}{4\pi}
  e^{i(\bm{k}'+\bm{k}'')\cdot\bm{r}}\,
  X^{l'l''}_{lm}(\hat{\bm{k}}',\hat{\bm{k}}'')
  \\
  = i^{l'+l''}
  \begin{pmatrix}
    l' & l'' & l \\ 0 & 0 & 0
  \end{pmatrix}
  j_{l'}(k'r) j_{l''}(k''r) C_{lm}(\hat{\bm{r}}),
  \label{eq:84}
\end{multline}
which can be derived from Eqs.~(\ref{eq:5}), (\ref{eq:13}),
(\ref{eq:42}), (\ref{eq:45}) and (\ref{eq:58}). Finally, for the
angular integration over $\bm{r}$, we only need Eqs.~(\ref{eq:13})
and (\ref{eq:42}).

Combining all the equations above, all the angular integrations in the
loop integral of Eq.~(\ref{eq:75}) are analytically evaluated.
Comparing the resulting expression with Eq.~(\ref{eq:57}), one can
read the corresponding component of the invariant power spectrum.
After straightforward calculations, the result is given by
\begin{multline}
  P^{l_1l_2;l}_{X_1X_2\,[2]}(k) =
  \frac{(-1)^l\{l\}}{2} \Pi^2(k)\,
  4\pi \int r^2 dr\,j_l(kr)
  \\ \times
  \sum_{\substack{l',l''\\l_1',l_1'',l_2',l_2''}}
  i^{l+l'+l''} (-1)^{l_2'+l_2''}
  \{l'\}\{l''\}
  \begin{pmatrix}
    l_1' & l_2' & l' \\ 0 & 0 & 0
  \end{pmatrix}
  \begin{pmatrix}
    l_1'' & l_2'' & l'' \\ 0 & 0 & 0
  \end{pmatrix}
  \\ \times
  \begin{pmatrix}
    l & l' & l'' \\ 0 & 0 & 0
  \end{pmatrix}
  \begin{Bmatrix}
    l_1 & l_2 & l \\
    l_1' & l_2' & l' \\
    l_1'' & l_2'' & l''
  \end{Bmatrix}
  \int \frac{k'^2dk'}{2\pi^2} \frac{k''^2dk''}{2\pi^2}
  P_\mathrm{L}(k') P_\mathrm{L}(k'')
  \\ \times
  j_{l'}(k'r) j_{l''}(k''r) 
  \hat{\Gamma}^{(2)\,l_1}_{X_1l_1'l_1''}(k',k'')
  \hat{\Gamma}^{(2)\,l_2}_{X_2l_2'l_2''}(k',k''),
  \label{eq:85}
\end{multline}
where the second-order invariant propagator is given by
Eq.~(\ref{eq:54}). The phase factor $i^{l+l'+l''}$ in the above
expression is real, because $l+l'+l''=\mathrm{even}$ due to the
special form of $3j$-symbol in front of the $6j$-symbol.

Besides the second-order renormalized bias function
$c^{(2)\,l}_{Xl_1l_2}(k_1,k_2)$, the two-dimensional integral of the
last term in Eq.~(\ref{eq:85}) is in fact given by a sum of products
of one-dimensional integrals, because the second-order propagator of
Eq.~(\ref{eq:54}) is given by a sum of terms in which the dependencies
on $k_1$ and $k_2$ are separated into factors in this case. After
numerically calculating the last integral as a function of $r$ and
storing the function as an interpolation table, the integral over $r$
is obtained by numerical integration. All the numerical integrations
we need are in the form of a one-dimensional Hankel transform, which
are evaluated by \textsc{FFTLog}. If the second-order renormalized
bias function is also given by a sum of terms in which the
dependencies on $k_1$ and $k_2$ are separated into factors, the whole
one-loop power spectrum, Eq.~(\ref{eq:85}), can be numerically
evaluated by a series of one-dimensional Hankel transforms with
\textsc{FFTLog}. This situation happens in the case of semilocal
models of bias, which concept is explained in detail in Sec.~VI of
Paper~I. More details of the situation are explained in
Secs.~\ref{subsec:OneSemiLocal} and \ref{sec:SimpleEx}.

The result of Eq.~(\ref{eq:85}) for the one-loop corrections can be
formally generalized to the cases of an arbitrary number of loop
corrections, provided that the higher-order propagators are given. The
details of the derivation and resulting expressions are given in the
Appendix.

\subsubsection{\label{subsubsec:PSred}
  Redshift space
}

The loop integral of Eq.~(\ref{eq:75}) can be similarly evaluated
even in redshift space. The difference between real space and redshift
space is that the second-order propagator is given by
Eq.~(\ref{eq:19}) in the latter, instead of Eq.~(\ref{eq:2}) in
the former. The product of tripolar spherical harmonics reduces to a
single tripolar spherical harmonics according to a formula
\cite{Khersonskii:1988krb,PaperI},
\begin{multline}
  X^{l_{z1}l_1'l_1''}_{L_1;l_1m_1}(\hat{\bm{z}},\hat{\bm{k}}',\hat{\bm{k}}'')
  X^{l_{z2}l_2'l_2''}_{L_2;l_2m_2}(\hat{\bm{z}},\hat{\bm{k}}',\hat{\bm{k}}'')
  =
  \sqrt{
    \{L_1\}\{L_2\}
  }
  \\ \times
  \sum_{l} (-1)^l \{l\}
  \left(l_1\,l_2\,l\right)_{m_1m_2}^{\phantom{m_1m_2}m}
  \sum_{l_z,l',l''} (-1)^{l_z+l'+l''}
  \{l_z\}\{l'\}\{l''\}
  \\ \times
  \begin{pmatrix}
    l_{z1} & l_{z2} & l_z \\
    0 & 0 & 0
  \end{pmatrix}
  \begin{pmatrix}
    l_1' & l_2' & l' \\
    0 & 0 & 0
  \end{pmatrix}
  \begin{pmatrix}
    l_1'' & l_2'' & l'' \\
    0 & 0 & 0
  \end{pmatrix}
  \sum_L \sqrt{\{L\}}  
  \\ \times
  \begin{Bmatrix}
    l_1 & l_2 & l \\
    l_{z1} & l_{z2} & l_z \\
    L_1 & L_2 & L
  \end{Bmatrix}
  \begin{Bmatrix}
    L_1 & L_2 & L \\
    l_1' & l_2' & l' \\
    l_1'' & l_2'' & l''
  \end{Bmatrix}
  X^{l_zl'l''}_{L;lm}(\hat{\bm{z}},\hat{\bm{k}}',\hat{\bm{k}}'').
  \label{eq:86}
\end{multline}
The rest of the calculations follow along the same line as in the case
of real space. The result for the invariant power spectrum is given by
\begin{multline}
  P^{l_1l_2;l\,l_z;L}_{X_1X_2\,[2]}(k) =
  \frac{(-1)^{l+l_z}\{l\} \{l_z\}\sqrt{\{L\}}}{2} \Pi^2(k)
  4\pi \int r^2 dr\,j_l(kr)
  \\ \times
  \sum_{\substack{l',l''\\l_1',l_1'',l_2',l_2''}}
  i^{l+l'+l''} (-1)^{l_2'+l_2''}
  \{l'\}\{l''\}
  \begin{pmatrix}
    l_1' & l_2' & l' \\ 0 & 0 & 0
  \end{pmatrix}
  \begin{pmatrix}
    l_1'' & l_2'' & l'' \\ 0 & 0 & 0
  \end{pmatrix}
  \\ \times
  \begin{pmatrix}
    l & l' & l'' \\ 0 & 0 & 0
  \end{pmatrix}
  \sum_{l_{z1},l_{z2},L_1,L_2}
  \sqrt{\{L_1\}\{L_2\}}
  \begin{pmatrix}
    l_{z1} & l_{z2} & l_z \\ 0 & 0 & 0
  \end{pmatrix}
  \\ \times
  \begin{Bmatrix}
    l_1 & l_2 & L \\
    l_{z1} & l_{z2} & l_z \\
    L_1 & L_2 & l
  \end{Bmatrix}
  \begin{Bmatrix}
    L_1 & L_2 & l \\
    l_1' & l_2' & l' \\
    l_1'' & l_2'' & l''
  \end{Bmatrix}
  \int \frac{k'^2dk'}{2\pi^2} \frac{k''^2dk''}{2\pi^2}
  \\ \times
  P_\mathrm{L}(k') P_\mathrm{L}(k'')
  j_{l'}(k'r) j_{l''}(k''r)
  \\ \times
  \hat{\Gamma}^{(2)\,l_1l_{z1};L_1}_{X_1l_1'l_1''}(k',k'';k,\mu)
  \hat{\Gamma}^{(2)\,l_2l_{z2};L_2}_{X_2l_2'l_2''}(k',k'';k,\mu),
  \label{eq:87}
\end{multline}
where the second-order invariant propagator is given by
Eq.~(\ref{eq:55}). Because of an identity of Eq.~(\ref{eq:81}) for the
$9j$-symbol, we readily see that Eq.~(\ref{eq:87}) in redshift space
reduces to Eq.~(\ref{eq:85}) in real space, when
$l_z = l_{z1} = l_{z2} = 0$.

The two-dimensional integral of the last factor in Eq.~(\ref{eq:87})
has the same structure as that in real space. Besides the second-order
renormalized bias function, the integrals reduce to a sum of
one-dimensional Hankel transforms which can be evaluated by
\textsc{FFTLog}. In semilocal models of bias, all of the integrals
reduce to essentially a series of one-dimensional Hankel transforms,
just as in the case of real space.

Also as in the case of real space, the result of Eq.~(\ref{eq:87})
for the one-loop corrections can be formally generalized to the cases
of an arbitrary number of loop corrections, provided that the
higher-order propagators are given. The details of the derivation and
resulting expressions are also given in the Appendix.

\subsection{\label{subsec:OneSemiLocal} Evaluations of one-loop
  integral in semilocal models of bias}

Besides the terms involving the second-order renormalized bias
function, all of the one-loop integrals appeared in the nonlinear
corrections of the power spectrum above reduce to a series of
one-dimensional Hankel transforms, as we see in the previous
subsection. In the case of semilocal models of bias, the second-order
renormalized bias function is also decomposed into a sum of products
of one-dimensional integrals. We see the situation more concretely
below.

\subsubsection{\label{subsubsec:SemiLocalGen}
  General considerations
}

The class of semilocal models of bias is defined in Paper I. In this
class of models, the tensor field
$F^\mathrm{L}_{i_1i_2\cdots}(\bm{q})$ at an arbitrary position
$\bm{q}$ in Lagrangian space is modeled by a (multivariate) function
of spatial derivatives of linear fields at the same position,
\begin{equation}
  \chi^{(a)}_{i_1i_2\cdots i_{L_a}}(\bm{q}) =
  \partial_{i_1}\partial_{i_2}\cdots\partial_{i_{L_a}}
  \psi^{(a)}(\bm{q}),
  \label{eq:88}
\end{equation}
where
\begin{equation}
  \psi^{(a)}(\bm{q}) =
  \int \frac{d^3k}{(2\pi)^3} e^{i\bm{k}\cdot\bm{q}}
  \delta_\mathrm{L}(\bm{k}) k^{-L_a} W^{(a)}(k)
  \label{eq:89}
\end{equation}
is the smoothed linear density field with an isotropic window function
$W^{(a)}(k)$ and $L_a$ is the rank of the linear tensor field of
Eq.~(\ref{eq:88}). The label ``$a$'' distinguishes different kinds
of the tensor field of a particular rank and window function, which
causally affect the tensor field $F_{Xlm}$.

In the semilocal models of bias, the biased tensor field
$F^\mathrm{L}_{i_1i_2\cdots}(\bm{q})$ at a given position in
Lagrangian space $\bm{q}$ is determined by a (multivariate) function
of $\chi^{(a)}_{i_1\cdots i_{L_a}}(\bm{q})$ at the same position.
Since the functional dependence should not explicitly depend on the
position $\bm{q}$ in the relation, one can specify the relation at the
position of coordinates origin, $\bm{q}=\bm{0}$, without loss of
generality. At this representative point, the linear tensor field of
Eq.~(\ref{eq:88}) is given by
\begin{equation}
  \chi^{(a)}_{i_1\cdots i_{L_a}} =
    i^{L_a}
    \int \frac{d^3k}{(2\pi)^3}\,
  \hat{k}_{i_1}\cdots\hat{k}_{i_{L_a}}
  \delta_\mathrm{L}(\bm{k}) W^{(a)}(k).
  \label{eq:90}
\end{equation}
The linear tensor fields are symmetric tensors by construction, and
can be decomposed into irreducible tensors $\chi^{(a)}_{lm}$ according
to the procedure explained in Sec.~VI of Paper~I. The resulting
irreducible tensor of the linear field is given by
\begin{equation}
  \chi^{(a)}_{lm}
  =
    i^{L_a}
  \int \frac{d^3k}{(2\pi)^3}\,
  \delta_\mathrm{L}(\bm{k})
  C_{lm}(\hat{\bm{k}}) W^{(a)}(k).
  \label{eq:91}
\end{equation}

In the semilocal models, the first-order and second-order
renormalized bias functions are given by (Sec.~IV~B~2 of Paper~I)
\begin{align}
    c^{(1)}_{Xl}(k)
  &
    = \sum_a b^{(1:a)}_{Xl} W^{(a)}(k),
  \label{eq:92}\\
    c^{(2)\,l}_{Xl_1l_2}(k_1,k_2)
  &
    = \sum_{a_1,a_2} b^{(2:a_1a_2)}_{Xl;l_1l_2}
    W^{(a_1)}(k_1) W^{(a_2)}(k_2),
  \label{eq:93}
\end{align}
where the scalar coefficients are defined by
\begin{align}
  \left\langle
  \frac{\partial F^\mathrm{L}_{Xlm}}{\partial\chi^{(a_1)}_{l_1m_1}}
  \right\rangle
  &=
    \frac{(-i)^{l+L_a}}{\sqrt{2l+1}}
    \delta_{ll_1} \delta_m^{m_1} b^{(1:a_1)}_{Xl},
    \label{eq:94}\\
  \left\langle
    \frac{\partial^2 F^\mathrm{L}_{Xlm}}
    {\partial\chi^{(a_1)}_{l_1m_1}\partial\chi^{(a_2)}_{l_2m_2}}
  \right\rangle
  &=
    \frac{i^{l-L_a-L_b}}{\sqrt{2l+1}}
    \left(l\,l_1\,l_2\right)_m^{\phantom{m}m_1m_2}
    b^{(2:a_1a_2)}_{Xl;l_1l_2}.
  \label{eq:95}
\end{align}
The bias parameters $b^{(1:a)}_{Xl}$ and $b^{(2:a_1a_2)}_{Xl;l_1l_2}$
are scalar constants, which are uniquely determined and calculated
when the model of the biased tensor field $F^\mathrm{L}_{Xlm}$ is
given by an explicit function of the linear tensor fields
$\chi^{(a)}_{lm}$'s. When the model of the biased tensor field is not
specified, the bias parameters can be considered free parameters which
are not determined only from rotational symmetry. Because of
Eq.~(\ref{eq:95}) and the symmetry of $3j$-symbols, the
second-order bias parameter satisfies an interchange symmetry,
\begin{equation}
  b^{(2:a_2a_1)}_{Xl;l_2l_1} = (-1)^{l+l_1+l_2}
  b^{(2:a_1a_2)}_{Xl;l_1l_2}.
  \label{eq:96}
\end{equation}

Both in real space and redshift space, the two-dimensional
integrals of the last terms in Eqs.~(\ref{eq:85}) and
(\ref{eq:87}) with second-order propagators of Eqs.~(\ref{eq:54})
and (\ref{eq:55}) involving the second-order renormalized bias
function are given by the following three types of integrals:
\begin{multline}
  \mathcal{A}^{l_1l_2}_{l'l'';l_1'l_1'';l_2'l_2''}(r)
  \equiv  \int \frac{k'^2dk'}{2\pi^2} \frac{k''^2dk''}{2\pi^2}
  P_\mathrm{L}(k') P_\mathrm{L}(k'')
  \\ \times
  j_{l'}(k'r) j_{l''}(k''r)
  c^{(2)\,l_1}_{X_1l_1'l_1''}(k',k'')
  c^{(2)\,l_2}_{X_2l_2'l_2''}(k',k''),
  \label{eq:97}
\end{multline}
\begin{multline}
  \mathcal{B}^{(n'n'')\,l_1l_2}_{l'l'';l_1'l_1''}(r) \equiv
  \int \frac{k'^2dk'}{2\pi^2} \frac{k''^2dk''}{2\pi^2}
  P_\mathrm{L}(k') P_\mathrm{L}(k'')
  \,(k')^{n'} (k'')^{n''}
  \\ \times
  j_{l'}(k'r) j_{l''}(k''r)
  c^{(2)\,l_1}_{X_1l_1'l_1''}(k',k'')
  c^{(1)}_{X_2l_2}(k'),
  \label{eq:98}
\end{multline}
and
\begin{multline}
  \mathcal{C}^{(n'n'')\,l_1}_{l'l'';l_1'l_1''}(r) \equiv
  \int \frac{k'^2dk'}{2\pi^2} \frac{k''^2dk''}{2\pi^2}
  P_\mathrm{L}(k') P_\mathrm{L}(k'')
  \,(k')^{n'} (k'')^{n''}
  \\ \times
  j_{l'}(k'r) j_{l''}(k''r)
  c^{(2)\,l_1}_{X_1l_1'l_1''}(k',k''),
  \label{eq:99}
\end{multline}
where integers $n',n''$ are one of $-1,0,1$. Substituting
Eqs.~(\ref{eq:92}) and (\ref{eq:93}) into
Eqs.~(\ref{eq:97})--(\ref{eq:99}), we straightforwardly have
\begin{align}
  \mathcal{A}^{l_1l_2}_{l'l'';l_1'l_1'';l_2'l_2''}(r)
  &= \sum_{\substack{a_1',a_1''\\a_2',a_2''}}
    b^{(2:a_1'a_1'')}_{X_1l_1;l_1'l_1''}
    b^{(2:a_2'a_2'')}_{X_2l_2;l_2'l_2''}
    \bar{\bar{\xi}}^{(a_1'a_2';0)}_{l'}(r)\,
    \bar{\bar{\xi}}^{(a_1''a_2'';0)}_{l''}(r),
  \label{eq:100}\\
  \mathcal{B}^{(n'n'')\,l_1l_2}_{l'l'';l_1'l_1''}(r)
  &= \sum_{a_1',a_1'',a_2}
    b^{(2:a_1'a_1'')}_{X_1l_1;l_1'l_1''}
    b^{(1:a_2)}_{X_2l_2}
    \bar{\bar{\xi}}^{(a_1'a_2;n')}_{l'}(r)\,
    \bar{\xi}^{(a_1'';n'')}_{l''}(r),
  \label{eq:101}\\
  \mathcal{C}^{(n'n'')\,l_1}_{l'l'';l_1'l_1''}(r)
  &= \sum_{a_1',a_1''}
    b^{(2:a_1'a_1'')}_{X_1l_1;l_1'l_1''}
    \bar{\xi}^{(a_1';n')}_{l'}(r)\,
    \bar{\xi}^{(a_1'';n'')}_{l''}(r),
  \label{eq:102}
\end{align}
where
\begin{align}
  \bar{\bar{\xi}}^{(ab;n)}_{l}(r)
  &= \int \frac{k^2dk}{2\pi^2} P_\mathrm{L}(k)
    \,k^n j_l(kr) W^{(a)}(k) W^{(b)}(k),
  \label{eq:103}\\
  \bar{\xi}^{(a;n)}_{l}(r)
  &= \int \frac{k^2dk}{2\pi^2} P_\mathrm{L}(k)
    \,k^n j_l(kr) W^{(a)}(k).
  \label{eq:104}
\end{align}
The functions of Eqs.~(\ref{eq:103}) and (\ref{eq:104}) reduce to
the function of Eq.~(\ref{eq:52}) in the limit of $R\rightarrow 0$,
and therefore essentially the same functions for large values of $r$.
These functions are cast into the Hankel transforms in the loop
corrections of the power spectrum, Eqs.~(\ref{eq:85}) and
(\ref{eq:87}), and thus the presence of the window function in the
integrand may not be important on large scales, $kR \ll 1$.
Other terms that do not involve second-order renormalized bias
function in the two-dimensional integrals of Eqs.~(\ref{eq:85}) and
(\ref{eq:87}) are similarly represented by the one-dimensional
integrals defined by Eqs.~(\ref{eq:103}), (\ref{eq:104}) and
(\ref{eq:52}). These integrals are all one-dimensional Hankel
transforms.

Similarly, the functions $Q^{(1)}_{1}(k)$, $Q^{(2)}_{ll'}(k)$ and
$R^{(n)}_{ll'}(k)$ of Eqs.~(\ref{eq:48}) and (\ref{eq:49}), which
appear in the loop corrections of the first-order propagators in
Eqs.~(\ref{eq:47}) and (\ref{eq:53}), are also represented by the same
types of integrals. Therefore, every integral in the one-loop
corrections of the power spectrum are represented by the integrals
Eqs.~(\ref{eq:103}), (\ref{eq:104}) and (\ref{eq:52}), following the
procedure described above. In practice, these functions,
$\xi^{(n)}_l(r)$, $\bar{\xi}^{(a;n)}_{l}(r)$ and
$\bar{\bar{\xi}}^{(a_1a_2;n)}_{l}(r)$, are numerically calculated and
stored in interpolation tables for possible indices and we can
evaluate the one-loop corrections of the power spectra,
Eqs.~(\ref{eq:85}) and (\ref{eq:87}), applying another Hankel
transform over the variable $r$. While the number of terms to evaluate
in this procedure can be large, all the numerical integrations are
one-dimensional ones which can be calculated by \textsc{FFTLog} in a
very short time.

\section{\label{sec:SimpleEx}
  Calculations with a simple example of tensor field}

\subsection{\label{subsubsec:SemiLocalExample} A simple example of
  bias through second-order derivatives of gravitational potential }

As a specific example of semilocal models of bias, we consider below
a simple class of bias models in which the biased tensor field
$F^\mathrm{L}_{Xlm}$ is a local function of only the second-order
derivatives of the linear potential field. The second-order
derivatives of the smoothed potential, with an appropriate
normalization, are given by
\begin{equation}
  \varphi_{ij}
  = \partial_i\partial_j \Laplace^{-1} \delta_R,
  \label{eq:105}
\end{equation}
where $\Laplace^{-1}$ is the inverse Laplacian, and $\delta_R(\bm{q})$
is a smoothed linear density contrast. At the representative point,
$\bm{q}=\bm{0}$, we have
\begin{equation}
  \delta_R =
  \int \frac{d^3k}{(2\pi)^3}\,
  \delta_\mathrm{L}(\bm{k}) W(kR),
  \label{eq:106}
\end{equation}
and $W(kR)$ is the window function in Fourier space with smoothing
radius $R$.

In this case, the second derivatives of the potential field with
negative sign, $-\varphi_{ij}$, correspond to a rank-2 linear tensor
field $\chi^{(a)}_{ij}$ of Eq.~(\ref{eq:90}) with an index $a$ fixed,
$L_a=2$ and $W^{(a)}(k) = W(kR)$. The irreducible components of the
potential derivatives are given by
\begin{align}
  \chi_{2m}
  &=
  -
  \int \frac{d^3k}{(2\pi)^3}
  \delta_\mathrm{L}(\bm{k})
    C_{2m}(\hat{\bm{k}})\,
    W(kR),
  \label{eq:107}\\
    \chi_{00}
  &=
      - \int \frac{d^3k}{(2\pi)^3}
    \delta_\mathrm{L}(\bm{k})
    C_{00}(\hat{\bm{k}})\,
    W(kR).
    \label{eq:108}
\end{align}
The scalar component just corresponds to $\chi_{00} = -\delta_R$.
Equations~(\ref{eq:107}) and (\ref{eq:108}) are simple examples of
Eq.~(\ref{eq:91}) in general situations.

The corresponding expressions of Eqs.~(\ref{eq:92}) and
(\ref{eq:93}) for the renormalized bias functions are given by
\begin{align}
  c^{(1)}_{Xl}(k)
  &= b^{(1)}_{Xl}W(kR), \quad
  \label{eq:109}\\
  c^{(2)\,l}_{Xl_1l_2}(k_1,k_2)
  &= b^{(2)}_{Xl;l_1l_2}
  W(k_1R) W(k_2R).
  \label{eq:110}
\end{align}
In the above, the integer $l$ of Eq.~(\ref{eq:109}) only take
values of $0,2$, and the integers $l_1,l_2$ of Eq.~(\ref{eq:110})
only takes values of $0,2$, and should satisfy a triangle inequality
$|l_1-l_2| \leq l \leq l_1+l_2$, as the $3j$-symbol in
Eq.~(\ref{eq:95}) indicates. Thus the number of possible parameters
$b^{(1)}_{Xl}$ and $b^{(2)}_{Xl;l_1l_2}$ is finite.

For the integrals of Eqs.~(\ref{eq:103}) and (\ref{eq:104}) in
this case, the field indices $a,b$ can be omitted, and we have
\begin{align}
  \bar{\bar{\xi}}^{(ab;n)}_l(r)
  &= \bar{\bar{\xi}}^{(n)}_l(r)
    \equiv
    \int \frac{k^2dk}{2\pi^2} P_\mathrm{L}(k)
    \,k^n j_l(kr) W^2(kR),
  \label{eq:111}\\
  \bar{\xi}^{(a;n)}_l(r)
  &= \bar{\xi}^{(n)}_l(r)
    \equiv
      \int \frac{k^2dk}{2\pi^2} P_\mathrm{L}(k)
  \,k^n j_l(kr) W(kR).
  \label{eq:112}
\end{align}
The functions of Eqs.~(\ref{eq:111}) and (\ref{eq:112}) reduce to
the function of Eq.~(\ref{eq:52}) in the limit of $R/r\rightarrow 0$,
and therefore essentially the same functions for large values of $r
\gg R$.

Substituting Eqs.~(\ref{eq:111}) and (\ref{eq:112}) into
Eqs.~(\ref{eq:100})--(\ref{eq:102}), we have
\begin{align}
  \mathcal{A}^{l_1l_2}_{l'l'';l_1'l_1'';l_2'l_2''}(r)
  &=
    b^{(2)}_{X_1l_1;l_1'l_1''}
    b^{(2)}_{X_2l_2;l_2'l_2''}
    \bar{\bar{\xi}}^{(0)}_{l'}(r)\,
    \bar{\bar{\xi}}^{(0)}_{l''}(r),
  \label{eq:113}\\
  \mathcal{B}^{(n'n'')\,l_1l_2}_{l'l'';l_1'l_1''}(r)
  &=
    b^{(2)}_{X_1l_1;l_1'l_1''}
    b^{(1)}_{X_2l_2}
    \bar{\bar{\xi}}^{(n')}_{l'}(r)\,
    \bar{\xi}^{(n'')}_{l''}(r),
  \label{eq:114}\\
  \mathcal{C}^{(n'n'')\,l_1}_{l'l'';l_1'l_1''}(r)
  &=
    b^{(2)}_{X_1l_1;l_1'l_1''}
    \bar{\xi}^{(n')}_{l'}(r)
    \bar{\xi}^{(n'')}_{l''}(r).
  \label{eq:115}
\end{align}
In the model we are considering here, the indices $l_1'$, $l_1''$,
$l_2'$, $l_2''$ only take values of $0,2$. Because the coefficients in
Eqs.~(\ref{eq:85}) and (\ref{eq:87}), the indices $l'$ and $l''$
satisfy triangle inequalities $|l_1' - l_2'| \leq l' \leq l_1' + l_2'$
and $|l_1'' - l_2''| \leq l'' \leq l_1'' + l_2''$, and also $l'$ and
$l''$ are both even numbers. Therefore, $l'$ and $l''$ take only
values of 0, 2, 4. In addition, the indices $l_1$ and $l_2$ satisfy
triangle inequalities $|l_1' - l_1''| \leq l_1 \leq l_1' + l_1''$ and
$|l_2' - l_2''| \leq l_2 \leq l_2' + l_2''$ because of
Eq.~(\ref{eq:95}). As noted in Paper I, depending on the parity of the
biased tensor $F_{Xlm}$, the integers $l_1+l_1'+l_1''$ and
$l_2+l_2'+l_2''$ should be even numbers if the biased tensor is a
normal tensor, and should be odd numbers if the biased tensor is a
pseudotensor. Therefore, $l_1$ and $l_2$ take only values of 0, 2, 4
for normal tensors and 1, 3 for pseudotensors in the one-loop power
spectra, otherwise the one-loop spectra are zero. The higher-rank
biased tensor fields are not generated only by the rank-2 linear
tensor fields in the one-loop order. Because of the above constraints,
the number of functions of Eqs.~(\ref{eq:113})--(\ref{eq:115}) are
finite and not too many.

We also need to evaluate one-loop integrals in the first-order
propagators $\hat{\Gamma}^{(1)}_X$ of Eqs.~(\ref{eq:47}) and
(\ref{eq:53}), which are needed to calculate Eqs.~(\ref{eq:78})
and (\ref{eq:79}). The integrals involving the renormalized bias
functions are given by Eqs.~(\ref{eq:48}) and (\ref{eq:49}). In
the model we are considering here, substituting Eqs.~(\ref{eq:109})
and (\ref{eq:110}) into Eqs.~(\ref{eq:48}) and (\ref{eq:49}),
we have
\begin{equation}
    Q^{(1)}_1(k) = Q^{(2)}_{ll'}(k) = 0, \quad
    R^{(n)}_{ll'}(k) = b^{(1)}_{Xl}\, \hat{R}^{(n)}_{l'}(k),
  \label{eq:116}
\end{equation}
where
\begin{equation}
    \hat{R}^{(n)}_{l}(k) \equiv
    k^{2-n} \int r\,dr\,j_{l}(kr)\,\bar{\xi}^{(n)}_{l}(r),
    \label{eq:117}
\end{equation}
and Eq.~(\ref{eq:112}) is used. The first two functions $Q^{(1)}_1(k)$
and $Q^{(2)}_{ll'}(k)$ are nonzero only when the linear tensor field
of rank-1, $\chi^{(a)}_{1m}$, contributes, and thus vanish in the
model we are considering here. Therefore, all the necessary loop
integrals to evaluate the one-loop power spectrum are given by a
series of one-dimensional Hankel transforms of
Eqs.~(\ref{eq:113})--(\ref{eq:116}), which can be calculated with the
\textsc{FFTLog} in a very short time.

\subsection{\label{subsec:ExampleIA} Calculation of
  the one-loop power spectrum and correlation function of rank-2
  tensor fields in a simple example}

We finally consider an example of numerical calculations of the
one-loop corrections to the power spectrum of a rank-2 tensor field,
which is related to the intrinsic alignment of the galaxy shapes. We
employ the simple model introduced in the previous subsection that the
rank-2 tensor field is locally determined by the second-order
derivatives of the gravitational potential in Lagrangian space,
Eq.~(\ref{eq:105}), and consider the autopower spectrum in real space.

Up to the one-loop approximation, the power spectrum of the rank-2
tensor field in real space is given by
\begin{equation}
  P^{22;l}_X(k) = P^{22;l}_{X\,[1]}(k) + P^{22;l}_{X\,[2]}(k),
  \label{eq:118}
\end{equation}
where
\begin{equation}
  P^{22;l}_{X\,[1]}(k)
  = \frac{\{l\}}{5}
    \begin{pmatrix}
      2 & 2 & l \\
      0 & 0 & 0
    \end{pmatrix}
    \left[\hat{\Gamma}^{(1)}_{X2}(k)\right]^2
    \Pi^2(k) P_\mathrm{L}(k),
  \label{eq:119}
\end{equation}
and 
\begin{multline}
  P^{22;l}_{X\,[2]}(k) = \frac{(-1)^l \{l\}}{2}
  \Pi^2(k)\,
  4\pi \int r^2 dr\,j_l(kr)
  \\ \times
  \sum_{\substack{l',l''\\l_1',l_1'',l_2',l_2''}}
  i^{l+l'+l''} (-1)^{l_2'+l_2''}
  \{l'\}\{l''\}
  \begin{pmatrix}
    l_1' & l_2' & l' \\ 0 & 0 & 0
  \end{pmatrix}
  \begin{pmatrix}
    l_1'' & l_2'' & l'' \\ 0 & 0 & 0
  \end{pmatrix}
  \\ \times
  \begin{pmatrix}
    l & l' & l'' \\ 0 & 0 & 0
  \end{pmatrix}
  \begin{Bmatrix}
    2 & 2 & l \\
    l_1' & l_2' & l' \\
    l_1'' & l_2'' & l''
  \end{Bmatrix}
  \int \frac{k'^2dk'}{2\pi^2} \frac{k''^2dk''}{2\pi^2}
  P_\mathrm{L}(k') P_\mathrm{L}(k'')
  \\ \times
  j_{l'}(k'r) j_{l''}(k''r) 
  \hat{\Gamma}^{(2)\,2}_{X_1l_1'l_1''}(k',k'')
  \hat{\Gamma}^{(2)\,2}_{X_2l_2'l_2''}(k',k''),
  \label{eq:120}
\end{multline}
as straightforwardly derived from Eqs.~(\ref{eq:78}) an
(\ref{eq:85}). The first-order propagator of rank-2 tensor,
$\hat{\Gamma}^{(1)}_{X2}(k)$, is given by Eq.~(\ref{eq:47}) with a
substitution of $l=2$. The number of summations over $l'$ in the
equation is finite because of the triangular inequality in the
$3j$-symbols. Substituting Eqs.~(\ref{eq:116}) and (\ref{eq:117})
into the equation, expanding the summation over $l'$, and substituting
numerical values of the $9j$-symbols, we explicitly derive
\begin{multline}
  \hat{\Gamma}^{(1)}_{X2}(k) =
  b^{(1)}_{X2}
  \Biggl\{
    W(kR)
    - \frac{2}{7}
    \left[
      \frac{1}{5} \hat{R}^{(0)}_0(k)
      - \frac{5}{7} \hat{R}^{(0)}_2(k)
      \right.
      \\
      \left.
      + \frac{18}{35} \hat{R}^{(0)}_4(k)
      + \frac{3}{35} \hat{R}^{(1)}_1(k)
      + \frac{1}{5} \hat{R}^{(1)}_3(k)
      - \frac{2}{7} \hat{R}^{(1)}_5(k)
    \right]
  \Biggr\},
  \label{eq:121}
\end{multline}
where $\hat{R}^{(n)}_{l}(k)$ is given by Eq.~(\ref{eq:117}). The
last integral is readily calculated by \textsc{FFTLog}, and the
function $P^{22;l}_{X\,[1]}(k)$ of Eq.~(\ref{eq:119}) is numerically
obtained.

The second-order propagator of rank-2 tensor,
$\hat{\Gamma}^{(2)\,2}_{Xl'l''}(k',k'')$, is given by
Eq.~(\ref{eq:54}) with substitution of $l=2$. All the nonzero
components are given by
\begin{align}
  \hat{\Gamma}^{(2)2}_{X00}(k_1,k_2)
  &= b^{(2)}_{X2;00} W(k_1R) W(k_2R),
  \label{eq:122}\\
  \hat{\Gamma}^{(2)2}_{X02}(k_1,k_2)
  &= b^{(2)}_{X2;02} W(k_1R) W(k_2R)
    + b^{(1)}_{X2}W(k_2R),
    \label{eq:123}\\
  \hat{\Gamma}^{(2)2}_{X20}(k_1,k_2)
  &= b^{(2)}_{X2;20} W(k_1R) W(k_2R)
    + b^{(1)}_{X2}W(k_1R),
  \label{eq:124}\\
  \hat{\Gamma}^{(2)2}_{X11}(k_1,k_2)
  &= \frac{\sqrt{6}}{5} b^{(1)}_{X2}
    \left[
    \frac{k_1}{k_2}W(k_1R) + \frac{k_2}{k_1}W(k_2R)
    \right],
  \label{eq:125}\\
  \hat{\Gamma}^{(2)2}_{X13}(k_1,k_2)
  &= - \frac{\sqrt{21}}{5}\, \frac{k_2}{k_1} b^{(1)}_{X2} W(k_2R),
  \label{eq:126}\\
  \hat{\Gamma}^{(2)2}_{X31}(k_1,k_2)
  &= - \frac{\sqrt{21}}{5}\, \frac{k_1}{k_2} b^{(1)}_{X2} W(k_1R),
  \label{eq:127}\\
  \hat{\Gamma}^{(2)2}_{X22}(k_1,k_2)
  &= b^{(2)}_{X2;22} W(k_1R) W(k_2R).
  \label{eq:128}
\end{align}
Substituting the above equations into Eq.~(\ref{eq:120}), the last
integral over $k'$ and $k''$ reduces to products of functions,
$\bar{\bar{\xi}}^{(n)}_l(r)$, $\bar{\xi}^{(n)}_l(r)$ and
$\xi^{(n)}_l(r)$. The integration over $r$ of these products is again
readily calculated by \textsc{FFTLog}. After summing over all the
possible integers $l'$, $l''$, $l_1'$, $l_1''$, $l_2'$, $l_2''$, the
number of which are finite, we numerically obtain the function
$P^{22;l}_{X\,[2]}(k)$ of Eq.~(\ref{eq:120}). Thus all the terms on
the lhs of Eq.~(\ref{eq:118}) are numerically evaluated by a series
of one-dimensional Hankel transforms in a very short time, thanks to
the \textsc{FFTLog}.

\begin{figure}
\centering
\includegraphics[width=20pc]{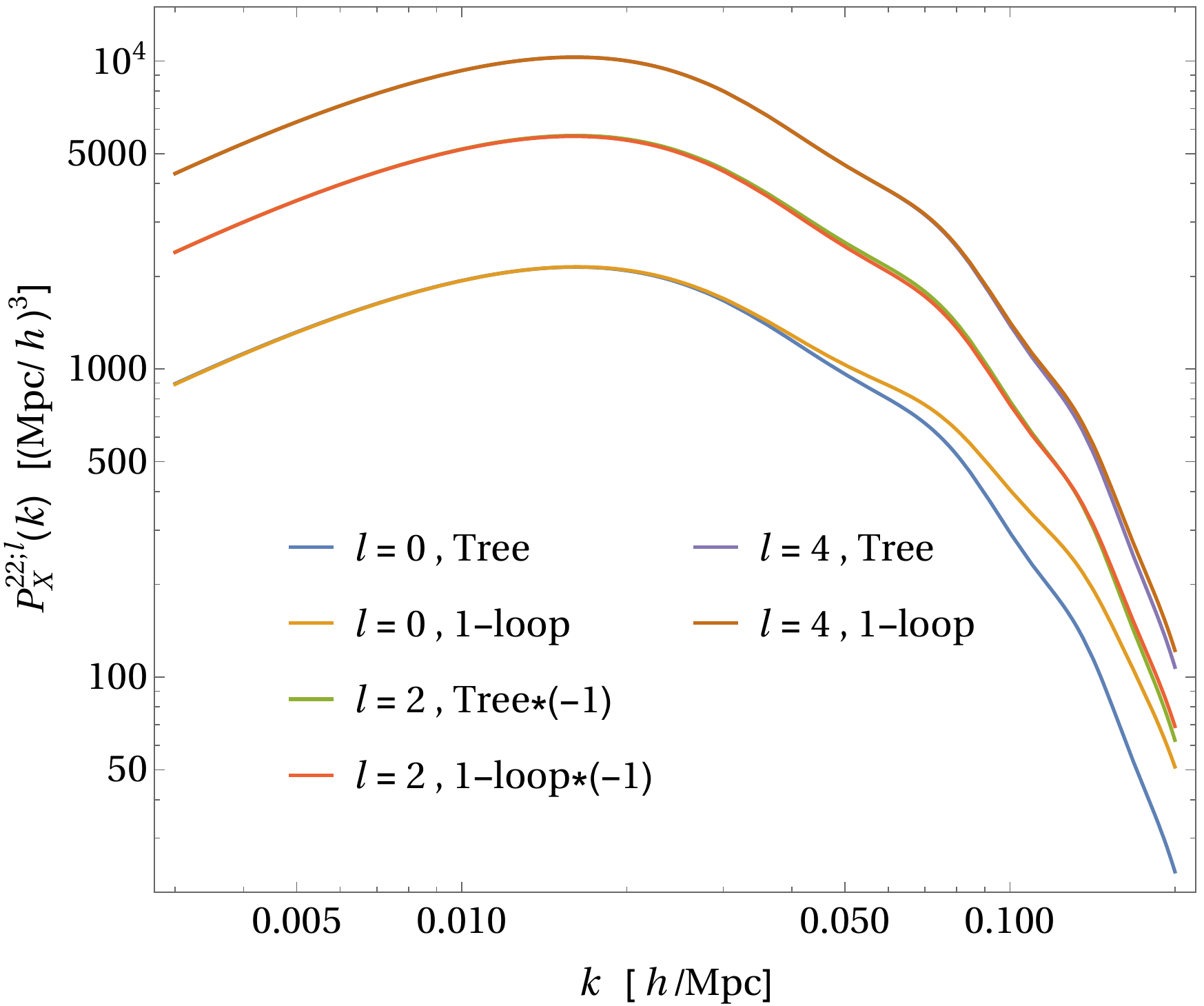}
\caption{\label{fig:1} The invariant power spectra $P^{22;l}_X(k)$ of
  a rank-2 tensor field in a simple model. The predictions of the tree
  (lowest-order) approximation and those of one-loop approximation are
  plotted, while they are almost overlapped with each other. From
  bottom to top, the power spectra with $l=0,2,4$ are shown
  respectively. The power spectrum with $l=2$ is negative and is
  multiplied by $-1$ in the plot. The absolute values of spectra with
  one-loop approximations are slightly larger than those of tree
  approximations, depending on the scales.}
\end{figure}
\begin{figure}
\centering
\includegraphics[width=20pc]{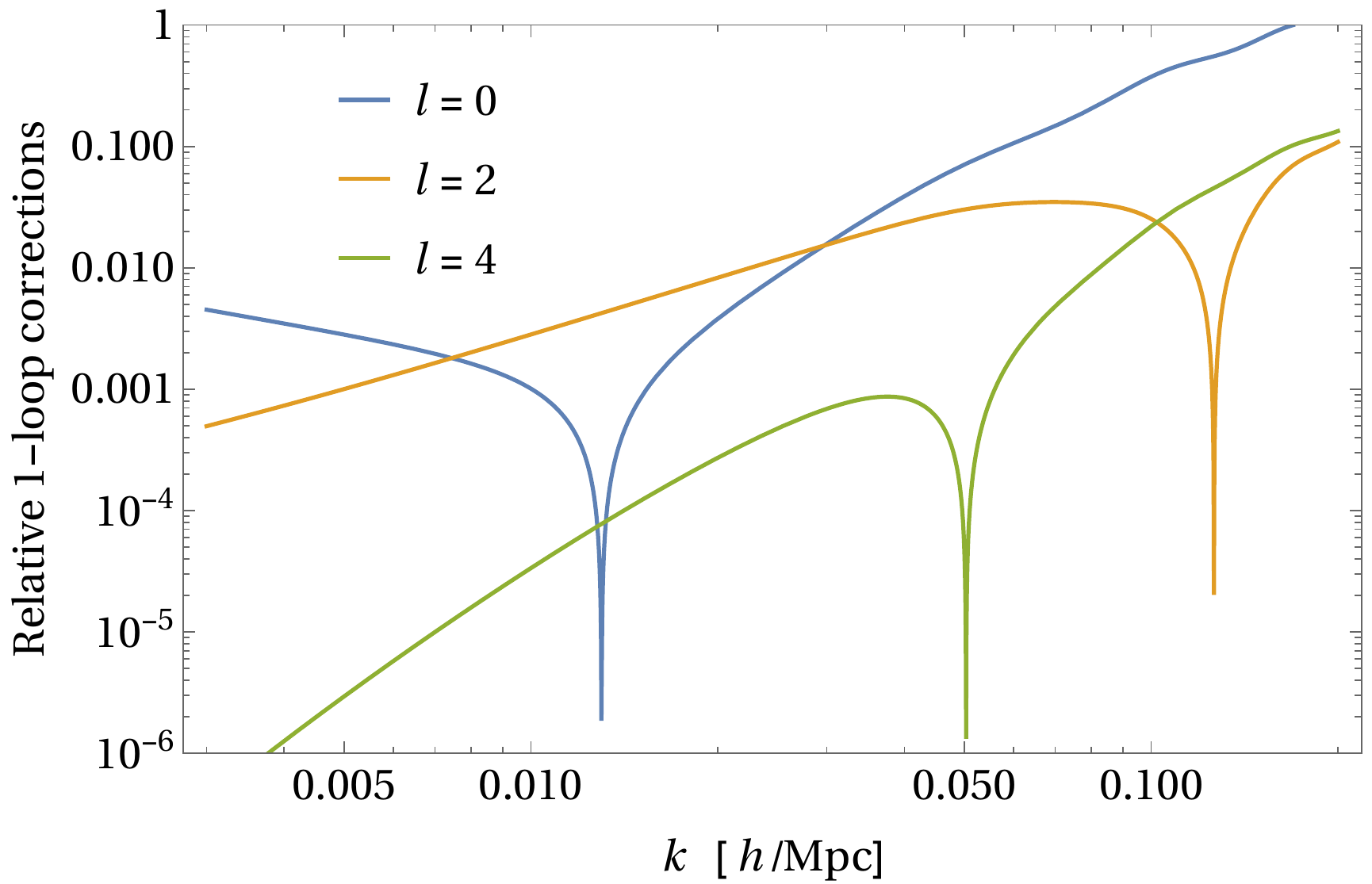}
\caption{\label{fig:2} Absolute values of contributions to the power
  spectra of one-loop corrections relative to those of lowest-order
  approximations, $|P^{22;l}_X(k)/P^{22;l}_{X,\mathrm{Tree}}(k)-1|$.}
\end{figure}

In the following, we show the results of numerical calculations of the
equations described above. The linear power spectrum $P_\mathrm{L}(k)$
of the mass density is calculated by a Boltzmann code \textsc{CLASS}
\cite{Lesgourgues:2011re,Blas:2011rf} with a flat $\Lambda$CDM model
and cosmological parameters $h=0.6732$,
$\Omega_{\mathrm{b}0}h^2=0.02238$, $\Omega_{\mathrm{cdm}}h^2=0.1201$,
$n_\mathrm{s}=0.9660$, and $\sigma_8=0.8120$ (Planck 2018
\cite{Planck:2018vyg}). We apply a smoothing radius of
$R=5\,h^{-1}\mathrm{Mpc}$ in the window function of the biasing model.
All of the bias parameters are simply substituted by unity,
$b^{(1)}_{X2} = b^{(2)}_{X2.00} = b^{(2)}_{X2;02} = b^{(2)}_{X2;20} =
b^{(2)}_{X2;22} = 1$, just for an illustrative purpose. The
interchange symmetry of Eq.~(\ref{eq:96}) requires an identity,
$b^{(2)}_{X2;02} = b^{(2)}_{X2;20}$. For the fast Hankel transforms,
we use a \textit{Mathematica} version of the numerical
code\footnote{\url{https://jila.colorado.edu/~ajsh/FFTLog/}}
\textsc{FFTLog} \cite{Hamilton:1999uv}. The adopted values of bias
parameters here are chosen just for illustrative purposes. The values
of bias parameters vary from sample to sample, depending on what kind
of objects are selected in a given survey. In particular, precise
values of second-order bias parameters have not been measured yet in
observations of intrinsic alignment, while the first-order parameters
are known to have orders of unity. However, as seen from
Eqs.~(\ref{eq:122})--(\ref{eq:128}), the nonlinear effects in
propagators are proportional to the bias parameters, and one can
roughly guess the effects of changing the values of bias parameters
from our simple choice of unity.

The results of the power spectra of the rank-2 tensor field,
$P^{22;l}_X(k)$, are shown in Fig.~\ref{fig:1}. The sign of the
spectrum $P^{22;2}_X(k)$ is negative. The predictions of the
lowest-order or tree approximation without loop corrections are
simultaneously shown in the plot as indicated by ``Tree'' in the
legends, together with the result including one-loop corrections as
indicated by ``1-loop'' in the legends. One immediately notices that
the effects of one-loop corrections are small on large scales for all
cases. In the case of $l=0$, the effect of one-loop corrections is
relatively large on smaller scales.

In Fig.~\ref{fig:2}, absolute values of relative fractions of the
one-loop corrections to the lowest-order approximations,
$|P^{22;l}_X(k)/P^{22;l}_{X,\mathrm{Tree}}(k)-1|$, are plotted, where
$P^{22;l}_{X,\mathrm{Tree}}(k)$ is the lowest-order prediction without
loop corrections. Overall, the larger the scales are (the smaller the
wavenumbers are), the smaller the effects of one-loop corrections are.
In the case of $l=0$, however, there remains a constant contribution
of the one-loop correction in the power spectrum, which is manifested
in Fig.~\ref{fig:2} that the difference from the linear theory are
increasing toward $k\rightarrow 0$ limit. In other cases of $l=2,4$,
this kind of shot noiselike contribution does not exist, because of
the asymptotic behavior of the spherical Bessel function
$j_l(z) \rightarrow 0$ when $z\rightarrow 0$ for $l\ge 1$. The shot
noise-like contribution in the one-loop power spectrum is already
known to exist also in the scalar perturbation theory, and a hint of
this effect is really seen in $N$-body simulations
\cite{Matsubara:2011ck,Sato:2011qr}. However, as shown below, this
constant contribution of the one-loop corrections does not
contribute to the correlation function at a finite separation, because
a constant in the power spectrum corresponds to the delta function at
the zero separation in the correlation function. One should note that the small
scales of $k \gtrsim 0.1\, h\,\mathrm{Mpc}^{-1}$ do not have physical
significance in the current situation, because they are smaller than
the smoothing radius $R = 5\, h^{-1}\mathrm{Mpc}$ in this example.
Therefore, the lowest-order approximation of the rank-2 tensor fields
with multipoles of $l=2,4$ is already accurate without loop
corrections, provided that our simplified assumption holds that they
are biased from only second derivatives of the linear gravitational
potential in the present model.

\begin{figure}
\centering
\includegraphics[width=20pc]{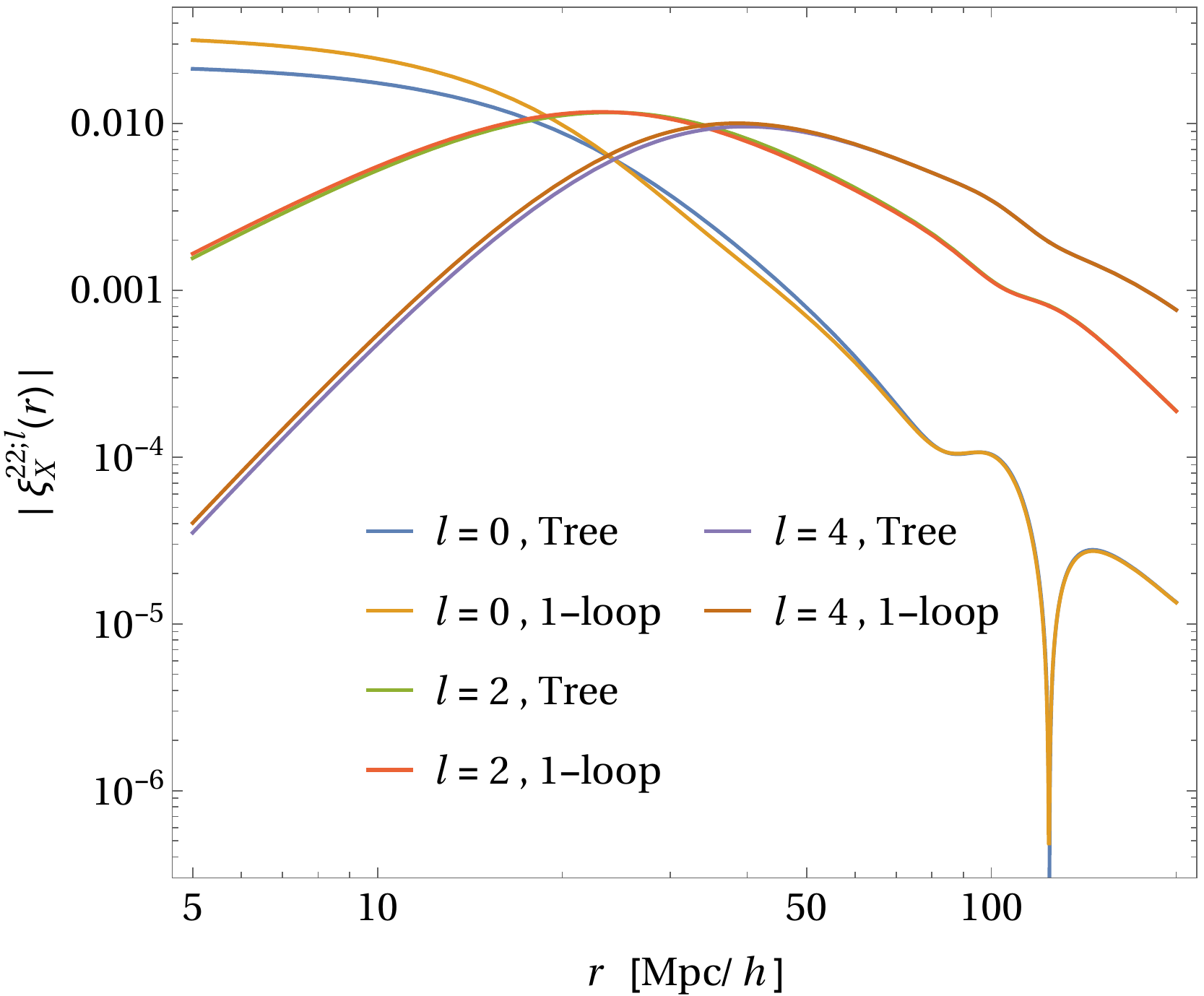}
\caption{\label{fig:3} The invariant correlation functions
  $\xi^{22;l}_X(r)$ of a rank-2 tensor field in a simple model. The
  predictions of tree (lowest-order) approximation and those of
  one-loop approximation for $l=0,2,4$ are plotted. The colors of
  lines correspond to those in Fig.~\ref{fig:1}. The absolute values
  of correlation functions with the one-loop approximation are larger
  than those of tree approximation on small scales,
  $r \lesssim 20\, h^{-1}\mathrm{Mpc}$. On scales of less
  than $100\,h^{-1}\mathrm{Mpc}$, the signs of the correlation
  functions are $\xi^{22;0}_X>0$, $\xi^{22;2}_X<0$ and
  $\xi^{22;4}_X>0$ in this example. The sign for $l=0$ changes beyond
  the BAO scale of $\sim 105\,h^{-1}\mathrm{Mpc}$. }
\end{figure}
\begin{figure}
\centering
\includegraphics[width=20pc]{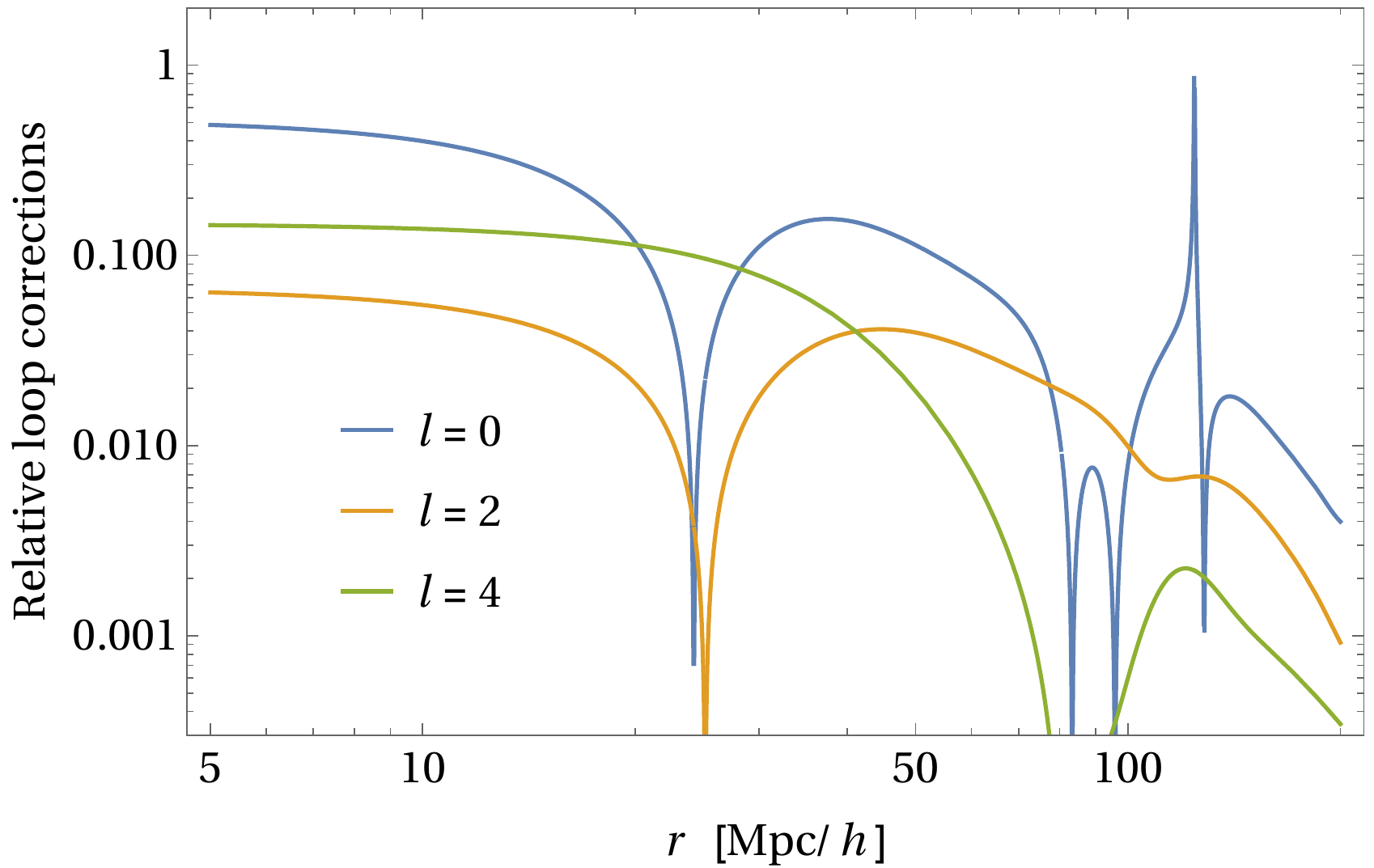}
\caption{\label{fig:4} Contributions to the correlation functions of
  one-loop corrections relative to those of lowest-order
  approximations, $|\xi^{22;l}_X(r)/\xi^{22;l}_{X,\mathrm{Tree}}(r)-1|$.
}
\end{figure}

We next calculate the correlation function of the rank-2 tensor field
$\xi^{22;l}_X(r)$. The invariant correlation function of the tensor
field is given by a Hankel transform of the invariant power spectrum
by Eq.~(\ref{eq:69}), which is again readily calculated by
\textsc{FFTLog}. The results are plotted in Fig.~\ref{fig:3}. The
absolute values of the relative fraction of the loop corrections,
$|\xi^{22;l}_X(r)/\xi^{22;l}_{X,\mathrm{Tree}}(r)-1|$, are plotted in
Fig.~\ref{fig:4}, where $\xi^{22;l}_{X,\mathrm{Tree}}(r)$ is the
lowest-order prediction without loop corrections. As seen in both
figures, the loop corrections to the correlation functions are not
negligible on scales of $r \lesssim 50\,h^{-1}\mathrm{Mpc}$ at the
level of 10\% order. On large scales of
$r \gtrsim 50\,h^{-1}\mathrm{Mpc}$, including the scales of BAO, the
lowest-order approximations without loop corrections are fairly
accurate roughly at the level of 1\% order or less, with an exception
near the scales of BAO, $\sim 100\,h^{-1}\mathrm{Mpc}$ in the case of
monopole component $l=0$, which crosses the zero value near there.
When the nonlinear bias parameters are different from our simple
choices of unity, the nonlinear effects in the figures scale according
to Eqs.~(\ref{eq:122})--(\ref{eq:128}). For example, when the
second-order bias parameters are smaller than unity, the magnitudes of
nonlinear effects reduce at least in proportion to these parameters.

In the above, we just exemplify applications of the formalism
developed in this paper to calculate loop corrections of the power
spectrum and correlation function, with a simple model that the tensor
bias is given only by second-order derivatives of the linear potential
of the gravitational field in Lagrangian space. We do not pursue
further interpretations of these particular results in detail, which
is beyond the scope of this paper. Some generalizations of the models
of shape bias, and detailed analyses of the particular predictions of
the rank-2 tensors in relation to the intrinsic alignment of galaxies
will be addressed in future work \cite{MATprep}.

\section{\label{sec:Conclusions}
  Conclusions
}

In this paper, we explicitly derive one-loop approximations of the
power spectrum and correlation function of tensor fields, using the
basic formulation developed in Paper I. The theory is built upon the
formalism of iPT and its generalization to include arbitrarily biased
tensor fields. As an example of applications, we explicitly and
numerically evaluate the one-loop power spectrum and correlation
function of the rank-2 tensor field with a model that the bias of the
tensor field is given by a function of second-order derivatives of the
gravitational potential. Higher-loop corrections are similarly
possible to calculate along the lines of this paper. Formal expressions
of the all-order power spectrum are also obtained in terms of the
higher-order propagators.

The original formalism of iPT is constructed in terms of Cartesian
wave vectors which appear in the functions of renormalized bias
functions and propagators. Since the tensor fields are decomposed into
spherical tensors in the present formalism, the functions which
involve wave vectors are naturally decomposed also into the spherical
basis. The renormalized bias functions and propagators are decomposed
into polypolar spherical harmonics in general. Unlike the previous
methods of the perturbation theory with spherical tensors, the
coordinates system of the spherical basis is not fixed to the direction
of a wave vector of perturbation mode, and the fully rotational
covariance is explicitly kept in the theory.

We define the rotationally invariant power spectrum and correlation
function of the tensor field on a spherical basis. As our formalism
uses the invariant functions for renormalized bias functions and
propagators, the invariant power spectrum and correlation functions
are also naturally represented by the invariant functions in loop
corrections. This is one of the benefits of the fully rotational
covariance of the theory in our formalism.

The redshift space distortions are also naturally incorporated into
formalism. Because of the rotational covariance of the theory, the
line of sight can be oriented in any direction. The renormalized bias
functions and propagators have additional dependencies on the
direction of the line of sight, which is also expanded by polypolar
spherical harmonics together with the directions of wave vectors. The
resulting propagators represented on a spherical basis are also
rotationally invariant. This is in contrast to the common treatment of
the redshift space distortions in perturbation theory, in which the
direction of the line of sight is fixed to, e.g., the third axis of
the coordinates system.

It has been known that the multidimensional integrations in the loop
corrections can be reduced to a series of one-dimensional Hankel
transforms by spherical decomposition of the wave vectors in the
kernel functions of the nonlinear perturbation theory and all the
necessary integrals can be evaluated by an algorithm of
\textsc{FFTLog}. The present formalism is also based on spherical
decomposition of the tensor field, essentially the same properties are
naturally derived. In fact, when we consider the simplest case of a
rank-0 tensor without bias, the previously known formula of the
perturbation theory using the FFT algorithm is exactly reproduced. In
the presence of (semi)local models of bias, the same technique can be
applied and all multidimensional integrations in the one-loop
corrections reduce to a series of one-dimensional Hankel transforms,
and thus are numerically calculated in a very short time using
\textsc{FFTLog}. In our rotationally covariant formalism with the
redshift space distortions, the stunning property of reducing the
dimensionality of integrals still holds even in redshift space.

In the last section, a simple example of the one-loop corrections of
the power spectrum and correlation function of the rank-2 tensor field
is presented and the results are numerically evaluated. In this
example, we assume the rank-2 tensor field is biased from the
second-order spatial derivatives of the gravitational potential. The
numerical integrations with \textsc{FFTLog} are quite stable and fast
enough. This example of tensor bias corresponds to the intrinsic
alignment driven by the mass density field and gravitational shear or
tidal field. Comparing these results with catalogs of halo shapes in
numerical simulations should be a straightforward and interesting
application of the present calculations.

More complicated modeling of the tensor bias, such as partial
inclusions of the property of the halo model, can be an interesting
application. Because of the generality of the present formalism, any
model of the tensor bias can be taken into account, as long as the
renormalized bias functions can be calculated. The bias model can even
be a singular function of the density field, as in the case of the
halo model
\cite{Matsubara:2012nc,Matsubara:2013ofa,Matsubara:2016wth}. This is
in contrast to other methods in which the bias function is expanded in
the Taylor series, and thus fully nonlinear or singular functions in
the bias relation, such as halo bias cannot be taken into account.
Investigations along this line are now in progress and will be
presented in future work \cite{MATprep}.

In this paper, we describe various techniques for calculating the loop
corrections to the power spectra and correlation functions, assuming
the Gaussian initial conditions. One can apply the same techniques to
calculate more general situations, some of which are described in
Paper~I, e.g., bispectra, trispectra, higher-order correlation
functions, and effects of (angle-dependent) primordial
non-Gaussianity, and so forth. The purpose of this paper is to develop
and provide techniques from a general point of view, and more specific
applications to the individual statistics which are more closely
related to observations are left for future work.

In realistic observations, most likely we can observe only projected
tensors on the two-dimensional sky, rather than three-dimensional
tensors themselves. It is technically straightforward to transform our
results of the power spectrum and correlation function into projected
tensors \cite{Vlah:2020ovg}. Characterizing galaxy shapes by
higher-rank tensors can also be an interesting direction of research
\cite{Kogai:2020vzz}. We will address the issue of projection effects
in Paper III \cite{PaperIII}. Relaxing the distant-observer
approximation, which is assumed in this paper and Papers~I and
III, generalizations to include full-sky and wide-angle effects will
be made in Paper~IV \cite{PaperIV}. Various methods for evaluating the
loop corrections described in this paper can be used directly in these
subsequent papers of the series as well.

\begin{acknowledgments}
  This work was supported by JSPS KAKENHI Grants No.~JP19K03835 and
  No.~21H03403.
\end{acknowledgments}

\newpage

\appendix
\onecolumngrid

\section{\label{app:PSall}
  Formal expressions of the nonlinear power spectrum to all orders
}

In the main text, we derive the full expressions of the one-loop power
spectra in real space and redshift space. Higher-loop corrections can
be calculated in the same way. The calculations of the higher-order
propagators with loop corrections are involved and tedious. However,
assuming the propagators are given, the expressions of the higher-loop
corrections, which generalize Eqs.~(\ref{eq:85}) and
(\ref{eq:87}), can be formally derived to all orders. The derivation
is simply a generalization of the derivation of Eqs.~(\ref{eq:85})
and (\ref{eq:87}). In this Appendix, the derivation is illustrated
and the resulting expressions are explicitly given.

We assume the Gaussian initial condition. The formal expression of the
power spectrum to all orders is formally given by straightforward
generalization of Ref.~\cite{Matsubara:2011ck} for scalar fields to
tensor fields,
\begin{multline}
  P^{(l_1l_2)}_{X_1X_2m_1m_2}(\bm{k})
  =
    i^{l_1+l_2}
  \Pi^2(\bm{k}) \sum_{n=1}^\infty \frac{1}{n!}
  \int \frac{d^3k^{(1)}}{(2\pi)^3}\cdots \frac{d^3k^{(n)}}{(2\pi)^3}
  (2\pi)^3\delta_\mathrm{D}^3(\bm{k}^{(1)}+\cdots +\bm{k}^{(n)} - \bm{k})
  \\ \times
  \hat{\Gamma}^{(n)}_{X_1l_1m_1}(\bm{k}^{(1)},\ldots,\bm{k}^{(n)})
  \hat{\Gamma}^{(n)}_{X_2l_2m_2}(-\bm{k}^{(1)},\ldots,-\bm{k}^{(n)})
  P_\mathrm{L}(k^{(1)}) \cdots P_\mathrm{L}(k^{(n)}).
  \label{eq:129}
\end{multline}
The expression of Eq.~(\ref{eq:129}) holds both in real space and in
redshift space. In redshift space, the power spectrum and propagators
depend also on the direction of the line of sight, $\hat{\bm{z}}$.

\subsection{\label{subsubsec:PSallReal}
  Real space
}

We first consider the power spectrum in real space. As shown in Paper
I, the directional dependence of $n$th-order normalized propagator on
wave vectors is expanded by the polypolar spherical harmonics as
\begin{equation}
  \hat{\Gamma}^{(n)}_{Xlm}(\bm{k}^{(1)},\cdots,\bm{k}^{(n)})
  =
  \sum_{\substack{l^{(1)},\ldots,l^{(n)}\\L^{(2)},\ldots,L^{(n-1)}}}
    \hat{\Gamma}^{(n)\,l;L^{(2)}\cdots L^{(n-1)}}_{Xl^{(1)}\cdots
      l^{(n)}}(k^{(1)},\cdots,k^{(n)})
    X^{l^{(1)}\cdots l^{(n)}}_{L^{(2)}\cdots L^{(n-1)};lm}
    (\hat{\bm{k}}^{(1)},\cdots,\hat{\bm{k}}^{(n)})
    \label{eq:130}
\end{equation}
The polypolar spherical harmonics of order $n$ are introduced in Paper~I,
and their definitions are
\begin{multline}
  X^{l_1l_2\cdots l_n}_{L_2L_3\cdots L_{n-1};lm}
  (\bm{n}_1,\bm{n}_2,\ldots,\bm{n}_n)
  = (-1)^{L_2+\cdots +L_{n-1}}
  \sqrt{\{L_2\}\{L_3\}\cdots\{L_{n-1}\}}
  \left(l\,l_1\,L_2\right)_{m}^{\phantom{m}m_1M_2}
  \left(L_2\,l_2\,L_3\right)_{M_2}^{\phantom{M_2}m_2M_3} \cdots
  \left(L_{n-2}\,l_{n-2}\,L_{n-1}\right)_{M_{n-2}}^{\phantom{M_{n-2}}m_{n-2}M_{n-1}}
  \\ \times
  \left(L_{n-1}\,l_{n-1}\,l_n\right)_{M_{n-1}}^{\phantom{M_{n-1}}m_{n-1}m_n}
  C_{l_1m_1}(\bm{n}_1) C_{l_2m_2}(\bm{n}_2) \cdots
  C_{l_nm_n}(\bm{n}_n),
  \label{eq:131}
\end{multline}
where $C_{lm}(\bm{n})$ are the spherical harmonics with Racah's
normalization defined in the main text by Eq.~(\ref{eq:3}), and
$(l_1\,l_2\,l_3)_{m_1}^{\phantom{m_1}m_2m_3}$ is our notation of the
$3j$-symbol in the main text by Eq.~(\ref{eq:6}). As in the main
text, we use a simplified notation for factors $\{L\} \equiv 2L+1$,
$\{l\} \equiv 2l+1$, and so forth. The polypolar spherical harmonics
are straightforward generalizations of the bipolar and tripolar
spherical harmonics, defined in the main text by Eqs.~(\ref{eq:5})
and (\ref{eq:20}), respectively.

Substituting Eq.~(\ref{eq:130}) into
Eq.~(\ref{eq:129}), a product of two polypolar spherical harmonics
appears. This product reduces to a single polypolar spherical
harmonics as (Paper I)
\begin{multline}
  X^{l_1l_2\cdots l_n}_{L_2L_3\cdots
    L_{n-1};lm}(\bm{n}_1,\bm{n}_2,\ldots,\bm{n}_n) 
  X^{l_1'l_2'\cdots l_n'}_{L_2'L_3'\cdots
    L_{n-1}';l'm'}(\bm{n}_1,\bm{n}_2,\ldots,\bm{n}_n) 
  =
  \sqrt{\{L_2\}\cdots\{L_{n-1}\}\{L_2'\}\cdots\{L_{n-1}'\}}
  \sum_{l''} (-1)^{l''}\{l''\}
  \left(l\,l'\,l''\right)_{mm'}^{\phantom{mm'}m''}
  \\ \times
  \sum_{l_1'',\ldots,l_n''} (-1)^{l_1''+\cdots+l_n''}
  \{l_1''\}\cdots\{l_n''\}
  \begin{pmatrix}
    l_1 & l_1' & l_1'' \\
    0 & 0 & 0
  \end{pmatrix}
  \cdots
  \begin{pmatrix}
    l_n & l_n' & l_n'' \\
    0 & 0 & 0
  \end{pmatrix}
  \sum_{L_2'',\ldots,L_{n-1}''} \sqrt{\{L_2''\}\cdots\{L_{n-1}''\}}
  \begin{Bmatrix}
    l & l' & l'' \\
    l_1 & l_1' & l_1'' \\
    L_2 & L_2' & L_2''
  \end{Bmatrix}
  \\ \times
  \begin{Bmatrix}
    L_2 & L_2' & L_2'' \\
    l_2 & l_2' & l_2'' \\
    L_3 & L_3' & L_3''
  \end{Bmatrix}
  \cdots
  \begin{Bmatrix}
    L_{n-2} & L_{n-2}' & L_{n-2}'' \\
    l_{n-2} & l_{n-2}' & l_{n-2}'' \\
    L_{n-1} & L_{n-1}' & L_{n-1}''
  \end{Bmatrix}
  \begin{Bmatrix}
    L_{n-1} & L_{n-1}' & L_{n-1}'' \\
    l_{n-1} & l_{n-1}' & l_{n-1}'' \\
    l_n & l_n' & l_n''
  \end{Bmatrix}
  X^{l_1''l_2''\cdots l_n''}_{L_2''L_3''\cdots
    L_{n-1}'';l''m''}(\bm{n}_1,\bm{n}_2,\ldots,\bm{n}_n).
  \label{eq:132}
\end{multline}
The integrals with a delta function in Eq.~(\ref{eq:129}) are
substituted by
\begin{equation}
  \int \frac{d^3k^{(1)}}{(2\pi)^3}\cdots \frac{d^3k^{(n)}}{(2\pi)^3}
  (2\pi)^3\delta_\mathrm{D}^3(\bm{k}^{(1)}+\cdots +\bm{k}^{(n)} -
  \bm{k}) \times \cdots
  = \int d^3r\, e^{-i\bm{k}\cdot\bm{r}} \int
  \frac{d^3k^{(1)}}{(2\pi)^3} \cdots \frac{d^3k^{(n)}}{(2\pi)^3}
  e^{i(\bm{k}^{(1)}+\cdots +\bm{k}^{(n)})\cdot\bm{r}}\times \cdots.
  \label{eq:133}
\end{equation}
Angular components of the wave vectors
$\bm{k}^{(1)},\ldots,\bm{k}^{(n)}$ in the integrand of
Eq.~(\ref{eq:129}) only appear in the polypolar spherical harmonics.
Consecutively using Eqs.~(\ref{eq:13}), (\ref{eq:58}),
(\ref{eq:42}) and (\ref{eq:45}), one can show that the necessary
integrals are given by
\begin{multline}
  \int
  \frac{d^2\hat{k}^{(1)}}{4\pi} \cdots \frac{d^2\hat{k}^{(n)}}{4\pi}
  e^{i(\bm{k}^{(1)}+\cdots +\bm{k}^{(n)})\cdot\bm{r}}
  X^{l_1l_2\cdots l_n}_{L_2L_3\cdots L_{n-1};lm}
  \left(\bm{k}^{(1)},\bm{k}^{(2)},\ldots,\bm{k}^{(n)}\right)
  = \frac{(-1)^l {\{l\}}}{(4\pi)^n}
  (-i)^{l_1+\cdots+l_n} (-1)^{L_2+\cdots+L_{n-1}}
  \{l_1\}\cdots\{l_n\}
  \\ \times
  j_{l_1}\left(k^{(1)}r\right) \cdots j_{l_n}\left(k^{(n)}r\right)
  \sqrt{\{L_2\}\cdots\{L_{n-1}\}}
  \begin{pmatrix}
    l & l_1 & L_2 \\ 0 & 0 & 0
  \end{pmatrix}
  \begin{pmatrix}
    L_2 & l_2 & L_3 \\ 0 & 0 & 0
  \end{pmatrix}
  \cdots
  \begin{pmatrix}
    L_{n-2} & l_{n-2} & L_{n-1} \\ 0 & 0 & 0
  \end{pmatrix}
  \begin{pmatrix}
    L_{n-1} & l_{n-1} & l_n \\ 0 & 0 & 0
  \end{pmatrix}
  C_{lm}(\bm{r}).
  \label{eq:134}
\end{multline}

Putting the above equations together, the angular integrations in
Eq.~(\ref{eq:129}) are analytically evaluated. Comparing the result
with Eq.~(\ref{eq:57}), or directly using Eq.~(\ref{eq:59}), we
finally derive the invariant power spectrum to all orders in real
space. The result is given by
\begin{multline}
  P^{l_1l_2;l}_{X_1X_2}(k)
  =(-1)^l\{l\}\,\Pi^2(k)\,
    4\pi \int r^2dr\, j_l(kr)
  \sum_{n=1}^\infty
  \frac{1}{n!}
  \sum_{
    \substack{
      l^{(1)},\ldots,l^{(n)},L^{(2)},\ldots,L^{(n-1)}\\
      l_1^{(1)},\ldots,l_1^{(n)},L_1^{(2)},\ldots,L_1^{(n-1)}\\
      l_2^{(1)},\ldots,l_2^{(n)},L_2^{(2)},\ldots,L_2^{(n-1)}
    }}
  i^{l+l^{(1)}+\cdots+l^{(n)}}
  (-1)^{l_2^{(1)}+\cdots+l_2^{(n)}}
  (-1)^{L^{(2)}+\cdots+L^{(n-1)}}
  \left\{l^{(1)}\right\}\cdots\left\{l^{(n)}\right\}
  \\ \times
  \left\{L^{(2)}\right\}\cdots\left\{L^{(n-1)}\right\}
  \sqrt{
    \left\{L_1^{(2)}\right\}\cdots\left\{L_1^{(n-1)}\right\}
    \left\{L_2^{(2)}\right\}\cdots\left\{L_2^{(n-1)}\right\}
  }
  \begin{pmatrix}
    l_1^{(1)} & l_2^{(1)} & l^{(1)} \\
    0 & 0 & 0
  \end{pmatrix}
  \cdots
  \begin{pmatrix}
    l_1^{(n)} & l_2^{(n)} & l^{(n)} \\
    0 & 0 & 0
  \end{pmatrix}
  \begin{pmatrix}
    l & l^{(1)} & L^{(2)} \\
    0 & 0 & 0
  \end{pmatrix}
  \\ \times
  \begin{pmatrix}
    L^{(2)} & l^{(2)} & L^{(3)} \\
    0 & 0 & 0
  \end{pmatrix}
  \cdots
  \begin{pmatrix}
    L^{(n-2)} & l^{(n-2)} & L^{(n-1)} \\
    0 & 0 & 0
  \end{pmatrix}
  \begin{pmatrix}
    L^{(n-1)} & l^{(n-1)} & l^{(n)} \\
    0 & 0 & 0
  \end{pmatrix}
  \begin{Bmatrix}
    l_1 & l_2 & l \\
    l_1^{(1)} & l_2^{(1)} & l^{(1)} \\
    L_1^{(2)} & L_2^{(2)} & L^{(2)}
  \end{Bmatrix}
  \\ \times
  \begin{Bmatrix}
    L_1^{(2)} & L_2^{(2)} & L^{(2)} \\
    l_1^{(2)} & l_2^{(2)} & l^{(2)} \\
    L_1^{(3)} & L_2^{(3)} & L^{(3)}
  \end{Bmatrix}
  \cdots
  \begin{Bmatrix}
    L_1^{(n-2)} & L_2^{(n-2)} & L^{(n-2)} \\
    l_1^{(n-2)} & l_2^{(n-2)} & l^{(n-2)} \\
    L_1^{(n-1)} & L_2^{(n-1)} & L^{(n-1)}
  \end{Bmatrix}
  \begin{Bmatrix}
    L_1^{(n-1)} & L_2^{(n-1)} & L^{(n-1)} \\
    l_1^{(n-1)} & l_2^{(n-1)} & l^{(n-1)} \\
    l_1^{(n)} & l_2^{(n)} & l^{(n)}
  \end{Bmatrix}
  \int \frac{{k^{(1)}}^2dk^{(1)}}{2\pi^2} \cdots
  \frac{{k^{(n)}}^2dk^{(n)}}{2\pi^2}
  P_\mathrm{L}\left(k^{(1)}\right) \cdots
  P_\mathrm{L}\left(k^{(n)}\right)
  \\ \times
  j_{l^{(1)}}\left(k^{(1)}r\right) \cdots
  j_{l^{(n)}}\left(k^{(n)}r\right)\
  \hat{\Gamma}^{(n)\,l_1;L_1^{(2)}\cdots
    L_1^{(n-1)}}_{X_1l_1^{(1)}\cdots
    l_1^{(n)}}\left(k^{(1)},\ldots,k^{(n)}\right) 
  \hat{\Gamma}^{(n)\,l_2;L_2^{(2)}\cdots
    L_2^{(n-1)}}_{X_2l_2^{(1)}\cdots
    l_2^{(n)}}\left(k^{(1)},\ldots,k^{(n)}\right).
  \label{eq:135}
\end{multline}
If the propagators in the spherical basis of the last two factors are
given by the sum of terms in which the dependencies on the
$k^{(1)},\ldots,k^{(n)}$ are separated, the above integrals are
calculated by a series of one-dimensional Hankel transforms using
\textsc{FFTLog}. For the gravitational evolution part, that is really
the case in the one-loop order as we explicitly show in the main text,
and also in the two-loop order \cite{Schmittfull:2016yqx}. For the
bias part, it is the case for the semilocal models as we see in the
main text.

\subsection{\label{subsubsec:PSallRed}
  Redshift space
}

A formal expression to all orders of the power spectrum in redshift
space can also be derived similarly as in the above case for real
space. The derivation is a straightforward generalization of the
one-loop expression given in Eq.~(\ref{eq:87}). All the necessary
equations are given in the above. We have extra dependence of the
propagators on the line-of-sight direction $\hat{\bm{z}}$. The
$n$th-order normalized propagator is expanded as (Paper I)
\begin{equation}
  \hat{\Gamma}^{(n)}_{Xlm}(\bm{k}_1,\cdots,\bm{k}_n;\hat{\bm{z}};k,\mu)
  =
  \sum_{\substack{l_z,l_1,\ldots,l_n\\L_1,\ldots,L_{n-1}}}
  \hat{\Gamma}^{(n)\,l\,l_z;L_1\cdots L_{n-1}}_{Xl_1\cdots l_n}
  (k_1,\cdots,k_n;k,\mu)
  X^{l_zl_1\cdots l_n}_{L_1\cdots L_{n-1};lm}
  (\hat{\bm{z}},\hat{\bm{k}}_1,\cdots,\hat{\bm{k}}_n).
  \label{eq:136}
\end{equation}
This expansion is substituted in Eq.~(\ref{eq:129}), and follows the
rest of the calculations in the case of real space above. The
consequent result is compared with Eq.~(\ref{eq:62}), or directly
substituted into Eq.~(\ref{eq:63}), and we finally derive the
invariant power spectrum to all orders in redshift space. The result
is given by
\begin{multline}
  P^{l_1l_2;l\,l_z;L}_{X_1X_2}(k,\mu)
  = (-)^{l+l_z} \{l\}\{l_z\} \sqrt{\{L\}}\, \Pi^2(k,\mu)\,
   4\pi \int r^2dr\, j_l(kr)
  \sum_{n=1}^\infty
  \frac{1}{n!}
  \sum_{
    \substack{
      l^{(1)},\ldots,l^{(n)},L^{(2)},\ldots,L^{(n-1)}\\
      l_1^{(1)},\ldots,l_1^{(n)},L_1^{(2)},\ldots,L_1^{(n-1)}\\
      l_2^{(1)},\ldots,l_2^{(n)},L_2^{(2)},\ldots,L_2^{(n-1)}
    }}
  i^{l+l^{(1)}+\cdots+l^{(n)}}
  (-1)^{l_2^{(1)}+\cdots+l_2^{(n)}}
  (-1)^{L^{(2)}+\cdots+L^{(n-1)}}
  \\ \times
  \left\{l^{(1)}\right\}\cdots\left\{l^{(n)}\right\}
  \left\{L^{(2)}\right\}\cdots\left\{L^{(n-1)}\right\}
  \sqrt{
    \left\{L_1^{(2)}\right\}\cdots\left\{L_1^{(n-1)}\right\}
    \left\{L_2^{(2)}\right\}\cdots\left\{L_2^{(n-1)}\right\}
  }
  \begin{pmatrix}
    l_1^{(1)} & l_2^{(1)} & l^{(1)} \\
    0 & 0 & 0
  \end{pmatrix}
  \cdots
  \begin{pmatrix}
    l_1^{(n)} & l_2^{(n)} & l^{(n)} \\
    0 & 0 & 0
  \end{pmatrix}
  \begin{pmatrix}
    l & l^{(1)} & L^{(2)} \\
    0 & 0 & 0
  \end{pmatrix}
  \\ \times
  \begin{pmatrix}
    L^{(2)} & l^{(2)} & L^{(3)} \\
    0 & 0 & 0
  \end{pmatrix}
  \cdots
  \begin{pmatrix}
    L^{(n-2)} & l^{(n-2)} & L^{(n-1)} \\
    0 & 0 & 0
  \end{pmatrix}
  \begin{pmatrix}
    L^{(n-1)} & l^{(n-1)} & l^{(n)} \\
    0 & 0 & 0
  \end{pmatrix}
  \sum_{l_{z1},l_{z2},L_1,L_2}
  \sqrt{\left\{L_1\right\}\left\{L_2\right\}}
  \begin{pmatrix}
    l_{z1} & l_{z2} & l_z \\
    0 & 0 & 0
  \end{pmatrix}
  \begin{Bmatrix}
    l_1 & l_2 & L \\
    l_{z1} & l_{z2} & l_z \\
    L_1 & L_2 & l
  \end{Bmatrix}
  \\ \times
  \begin{Bmatrix}
    L_1 & L_2 & l \\
    l_1^{(1)} & l_2^{(1)} & l^{(1)} \\
    L_1^{(2)} & L_2^{(2)} & L^{(2)}
  \end{Bmatrix}
  \begin{Bmatrix}
    L_1^{(2)} & L_2^{(2)} & L^{(2)} \\
    l_1^{(2)} & l_2^{(2)} & l^{(2)} \\
    L_1^{(3)} & L_2^{(3)} & L^{(3)}
  \end{Bmatrix}
  \cdots
  \begin{Bmatrix}
    L_1^{(n-2)} & L_2^{(n-2)} & L^{(n-2)} \\
    l_1^{(n-2)} & l_2^{(n-2)} & l^{(n-2)} \\
    L_1^{(n-1)} & L_2^{(n-1)} & L^{(n-1)}
  \end{Bmatrix}
  \begin{Bmatrix}
    L_1^{(n-1)} & L_2^{(n-1)} & L^{(n-1)} \\
    l_1^{(n-1)} & l_2^{(n-1)} & l^{(n-1)} \\
    l_1^{(n)} & l_2^{(n)} & l^{(n)}
  \end{Bmatrix}
  \\ \times
  \int \frac{{k^{(1)}}^2dk^{(1)}}{2\pi^2} \cdots
  \frac{{k^{(n)}}^2dk^{(n)}}{2\pi^2}
  P_\mathrm{L}\left(k^{(1)}\right) \cdots
  P_\mathrm{L}\left(k^{(n)}\right)
  j_{l^{(1)}}\left(k^{(1)}r\right) \cdots
  j_{l^{(n)}}\left(k^{(n)}r\right)\
  \\ \times
  \hat{\Gamma}^{(n)\,l_1l_{z1};L_1L_1^{(2)}\cdots
    L_1^{(n-1)}}_{X_1l_1^{(1)}\cdots
    l_1^{(n)}}\left(k^{(1)},\ldots,k^{(n)};k,\mu\right) 
  \hat{\Gamma}^{(n)\,l_2l_{z2};L_2L_2^{(2)}\cdots
    L_2^{(n-1)}}_{X_2l_2^{(1)}\cdots
    l_2^{(n)}}\left(k^{(1)},\ldots,k^{(n)};k,\mu\right).
  \label{eq:137}
\end{multline}
As a consistency check, we readily see that this expression exactly
reduces to Eq.~(\ref{eq:135}) when we substitute
$l_z=l_{z1}=l_{z2}=0$, using Eq.~(\ref{eq:81}). In the case of
semilocal models of bias, the radial integrals are calculated by
a series of one-dimensional Hankel transforms using \textsc{FFTLog},
just as in the case of real space.


\twocolumngrid
\renewcommand{\apj}{Astrophys.~J. }
\newcommand{\aap}{Astron.~Astrophys. }
\newcommand{\aj}{Astron.~J. }
\newcommand{\apjl}{Astrophys.~J.~Lett. }
\newcommand{\apjs}{Astrophys.~J.~Suppl.~Ser. }
\newcommand{\apss}{Astrophys.~Space Sci. }
\newcommand{\cqg}{Class.~Quant.~Grav. }
\newcommand{\jcap}{J.~Cosmol.~Astropart.~Phys. }
\newcommand{\mnras}{Mon.~Not.~R.~Astron.~Soc. }
\newcommand{\mpla}{Mod.~Phys.~Lett.~A }
\newcommand{\pasj}{Publ.~Astron.~Soc.~Japan }
\newcommand{\physrep}{Phys.~Rep. }
\newcommand{\ptp}{Progr.~Theor.~Phys. }
\newcommand{\ptep}{Prog.~Theor.~Exp.~Phys. }
\newcommand{\jetp}{JETP }
\newcommand{\jhep}{Journal of High Energy Physics}


\end{document}